\documentclass[article]{aa}

\usepackage{graphicx}
\usepackage{natbib}
\usepackage{amssymb}
\usepackage[figuresright]{rotating}

\begin{document}

\newcommand{\gapprox}{$\stackrel {>}{_{\sim}}$}   %greater/less approx.
\newcommand{\lapprox}{$\stackrel {<}{_{\sim}}$}

\title{Near- and Far-Infrared  Counterparts of Millimeter Dust Cores\\ in the Vela Molecular Ridge Cloud D
\thanks{Based on observations collected at NTT and 3.6m telescope
(ESO - La Silla, Chile).}}

   \author{M. De Luca
          \inst{1,2},
          %\and
          T. Giannini
            \inst{1}, 
          D. Lorenzetti
            \inst{1},
          F. Massi
            \inst{3},
	  D. Elia
     	    \inst{4},	    
            \and
          B. Nisini
            \inst{1}
           }
   \offprints{M. De Luca}

   \institute{INAF - Osservatorio Astronomico di Roma,
              via Frascati 33, 00040 Monte Porzio Catone (Roma), Italy\\
              \email{deluca, giannini, lorenzetti, nisini@oa-roma.inaf.it}
              \and
              Dipartimento di Fisica - Universit\`{a} di Roma ``Tor Vergata",
              via della Ricerca Scientifica 1, 00033 Roma, Italy
              \and
              INAF - Osservatorio Astrofisico di Arcetri,
              Largo E.Fermi 5, 50125 Firenze, Italy\\
              \email{fmassi@arcetri.astro.it} 
              \and 
              Dipartimento di Fisica - Universit\`{a} di Lecce, CP 193, 73100 Lecce, Italy
             }

  \abstract
  {}% context heading (optional)
  {Identify the young protostellar counterparts associated to dust millimeter cores of the Vela Molecular Ridge Cloud D through new IR observations ($H_2$ narrow-band at 2.12\,$\mu$m and $N$ broad band at 10.4\,$\mu$m) along with an investigation performed on the existing IR catalogues.}
  {The association of $mm$ continuum emission with infrared sources from catalogues (IRAS, MSX, 2MASS), $JHK$ data from the literature and new observations, has been established according to spatial coincidence, infrared colours and spectral energy distributions.} 
  {Only 7 out of 29 resolved $mm$ cores (and 16 out of the 26 unresolved ones) do not exhibit signposts of star formation activity. The other ones are clearly associated with: far-IR sources, $H_2$ jets pointing back to embedded objects not (yet) detected or near-IR objects showing a high intrinsic colour excess. The distribution of the spectral indices pertaining to the associated sources is peaked at values typical of Class I objects, while three objects are signalled as candidates Class\,0 sources. Objects with far-IR colours similar to those of T-Tauri and Herbig Ae/Be stars seem to be very few. An additional population of young objects exists associated not with the $mm$-cores, but with both the diffuse warm dust emission and the gas filaments. We remark the high detection rate (30\%) of $H_2$ jets driven by sources located inside the $mm$-cores. They appear not driven by the most luminous objects in the field, but rather by less luminous objects in young clusters, testifying the co-existence of both low- and intermediate-mass star formation.} 
  {The presented results reliably describe the young population of VMR-D. However, the statistical evaluation of activity vs inactivity of the investigated cores, even in good agreement with results found for other star forming regions, seems to reflect the limiting sensitivity of the available facilities rather than any property intrinsic to the $mm$-condensations.} 
  
\keywords{Stars: formation --  Infrared: stars -- ISM: individual objects: Vela Molecular Ridge -- ISM: clouds -- ISM: jets and outflows -- Catalogs}

\authorrunning{M. De Luca et al.}
\titlerunning{IR Counterparts of $mm$ Dust Cores in the VMR-D}

\maketitle

\section{Introduction}

The association of far-infrared (FIR) sources with the gas and dust emission cores in  Giant Molecular Clouds (GMCs) is the starting point of any effort aimed to observationally study high- and intermediate-mass star formation modalities. Indeed the Galactic matter is distributed in such a way that prevents to have a significant number of GMCs located near our Sun. Also low-mass star formation occurs in GMCs, but GMCs are usually found at distances greater than 1-2\,Kpc, where the sensitivity of the current fore-front instrumentation makes it possible to sample only the most luminous (i.e. the most massive) objects. As a consequence, studying low-mass star formation is limited to nearby dark clouds, leaving GMCs as the privileged targets for high-mass studies. Exceptions exist, which are represented by the Orion and Vela GMCs (usually named Vela Molecular Ridge - VMR) and by the OB association in Scorpius-Centaurus and Perseus, whose distances range between 300 and 700\,pc. Orion GMC is by far the most studied star formation site at all wavelengths and both modalities (high- and low-mass) are observationally documented to take place in it (e.g. Chen \& Tokunaga 1994). Its location outside the plane of the Galaxy ($b$\,$\approx$\,-20$^{\circ}$) has originally sanctioned its leading r\^{o}le in star formation studies, when extinction and crowding represented insurmountable observational barriers. Also Sco-Cen and Per OB associations are located at about +20$^{\circ}$ and -20$^{\circ}$, respectively. The rapid growth of multi-frequency facilities at an increasing spatial resolution, makes it possible now to study, with enough accuracy, even extincted and crowded regions, as the VMR, in the plane of the Galaxy. Thanks to the forthcoming facilities, investigating the plane will be easier and easier and the properties derivable from the VMR will be the most suitable for obtaining a direct comparison with those of the other galaxies, whose planes represent the main volume we are able to sample.\\
The VMR is a complex of four GMCs (Murphy \& May 1991; Yamaguchi et al. 1999) and it is probably one of the best regions for studying the processes involved in star formation (clustering, isolation, matter outflows). It is located in the galactic plane ($b\,=\,\pm\,3^{\circ}$) outside the solar circle ($\ell\,\sim\,260^{\circ}\,-\,275^{\circ}$) and most of the gas is at a distance of $\sim$\,700\,pc (Liseau et al. 1992). So far we have accumulated a large data-base on VMR clouds through ground-based observations from near-IR (NIR) to $mm$ wavelengths. In particular, our analysis, based on the IRAS point source catalog \cite{IRAS-PSC1988}, unveiled a remarkable concentration of red FIR sources with bolometric luminosities $<$\,$10^4\,L_{\odot}$ (Liseau et al. 1992, Lorenzetti et al. 1993). These are very young intermediate-mass stars and we found (Massi et al. 2000, 2003) that FIR sources with $L_{bol}\,>\,10^3\,L_{\odot}$ coincide with young embedded clusters (size $\sim$\,0.1\,pc, $\sim$\,50-100 members). In particular, we find that the region of the VMR named {\it cloud D} (hereinafter VMR-D) hosts a large number of these, revealing a high efficiency in this mode of star formation. At the same time, the presence of IRAS sources having bolometric luminosities of only few solar luminosities, shows that the formation of isolated, low-mass stars is also active in this region.\\
We have searched the region around a complete sample of IRAS sources in VMR-D for protostellar jets, using NIR imaging and spectroscopy (Lorenzetti et al. 2002), and we have discovered a significant number of shock tracers ($H_2$ and $[FeII]$ line emission knots), which signal the presence of protostellar jets in VMR-D. We have also clarified the details of the interactions between jets and circumstellar environment as well as the properties of the exciting sources (Caratti o Garatti et al. 2004, Giannini et al. 2001, 2005).\\
Recently we mapped a $\sim$\,1\,deg$^2$ area of the VMR-D in the 1.2\,mm continuum of dust emission, along with the $^{12}$CO(1--0) and $^{13}$CO(2--1) transitions (see Fig.\,1 in Massi et al. 2007 and Figs.\,3-4 in Elia et al. 2007). The aim of the present paper is to correlate dust map and available FIR and NIR catalogues to have a complete census of the VMR-D young population that allows us to assess the pre-main sequence evolutionary stage properties. The correspondence between dust cores and molecular clumps has been already presented in Elia et al. (2007).\\
An increasing number of millimeter surveys towards star formation regions along with a comparison with existing FIR catalogues is now available. For the VMR complex, in particular, an investigation of the Vela-C cloud has been carried out by Moriguchi et al. (2003) and by Baba et al. (2006), and $mm$-studies of few individual sources belonging to VMR-D are reported by Faundez et al. (2004) and Fontani et al. (2005).\\
In the following, we give in Sect.\,\ref{par:The-investigated-region} a short summary of the results on the $mm$-maps (both dust and gas) we have presented elsewhere. The criteria for associating IR counterparts to the dust emission cores are discussed in Sect.\,\ref{par:Association-with-point-sources-catalogues}, along with the analysis of a particular selected area, as an example. New IR observations are also presented in this section. The results are then summarized in Sect.\,\ref{par:Results} and discussed in Sect.\,\ref{par:Discussion}. In appendix A, the associations for all the regions of dust emission are discussed separately.

\section{The investigated region} \label{par:The-investigated-region}
Our observations of the VMR-D carried out in the millimeter range with the SEST (ESO - La Silla) telescope are described elsewhere and consist of three maps of about 1\,deg$^2$, both in the dust continuum emission at 1.2\,mm \cite{Massi2007} and in the molecular transitions $^{12}$CO(1-0), at 2.6\,mm, and $^{13}$CO(2-1), at 1.3\,mm (Elia et al. 2007).\\
The $^{12}$CO data (resolution: 43", sampling: 50") outline a filamentary distribution of diffuse molecular emission connecting regions of enhanced intensity where there is evidence of clustered, intermediate-mass, star formation in progress (Elia et al. 2007, Massi et al. 2007). The map of $^{13}$CO emission, because of its lower abundance with respect to $^{12}$CO, traces denser regions of the molecular cloud, which present a quite clumpy structure. Summarizing the results from the $^{13}$CO map, we have found 49 clumps with mass, size and mean velocity ranging from 2 to 140\,$M_{\odot}$, from 0.15 to 0.67\,pc and from 1 to 13\,km\,s$^{-1}$, respectively (see Tab.\,3 in Elia et al. 2007).\\
The dust map (resolution: 24", corresponding to about 0.1\,pc, in fast scanning mode of 80\,arcsec\,s$^{-1}$) also shows a clumpy structure and we have individuated 29 cores of mass and size in the ranges 0.2 - 80\,$M_{\odot}$ and 0.03 - 0.3\,pc respectively (see Tab.\,1 in Massi et al. 2007), almost all of them nearly coincident with the brightest regions of the velocity integrated CO maps.
In addition, other 26 cores, whose size is under the spatial resolution, have been identified, even if their genuine nature remains uncertain.

\section{Association with point-sources} \label{par:Association-with-point-sources-catalogues}
In this work we aim to correlate the millimeter emission with objects from both new IR images (1-10\,$\mu$m) carried out on selected areas (see Sect.\,\ref{par:Observations}) and the infrared existing catalogs of point sources in order to find out the sources that dominate the dust cores heating. In particular, the considered catalogs are: {\it (i)} the IRAS point source catalog \cite{IRAS-PSC1988}, that provides flux densities at 12, 25, 60 and 100\,$\mu$m; {\it (ii)} the MSX (Midcourse Space Experiment) point source catalog \cite{Price2001} at 8.28, 12.13, 14.65 and 21.3\,$\mu$m; {\it (iii)} the 2MASS point source catalog \cite{Cutri2003}, based on $J$, $H$, $Ks$ bands, complemented with literature data \cite{Massi1999} of deeper IRAC2 images \cite{Moorwood1992}.

\subsection{Observations} \label{par:Observations}
An observational campaign has been carried out by using NIR (narrow-band $H_2$\,1-0S(1) at $\lambda$\,=\,2.13\,$\mu$m) and mid-IR (MIR) ($N$ broad-band) imaging facilities to observe fields selected from the $mm$-emission maps. All the recognized dust cores (except one) have been imaged in $H_2$, with the aim of searching for protostellar jet evidences.  The 10.4\,$\mu$m survey covered those cores associated to the presence of a young embedded cluster (Massi et al. 2003, 2006), aiming, thanks to an adequate spatial resolution, to pick-up those source(s) (if any) that dominate(s) the detected fluxes at FIR wavelengths.

\subsubsection{NIR imaging}
Broadband $J$, $H$, $Ks$ and narrow-band images in the $H_2$ 1-0S(1) ($\lambda$\,=\,2.13\,$\mu$m, $\Delta \lambda$\,=\,0.03\,$\mu$m) were obtained in January 2006 with SofI\footnote{$J$ and $H$ images were obtained only for those fields containing the young clusters.} (Lidman et al. 2000) at ESO-NTT (La Silla, Chile). The total field of view is 4.9$\times$4.9\,arcmin$^2$, which corresponds to a plate scale of 0.29\,arcsec/pixel. All the observations were obtained by dithering the telescope around the pointed position and the raw imaging data were reduced by using standard procedures for bad pixel removal, flat fielding, and sky subtraction. 

\subsubsection{N-band imaging}
Imaging in the N10.4 broadband filter was carried out in January 2006 with Timmi2 (Saviane \& Doublier 2005) at the 3.6\,m ESO telescope (La Silla, Chile). The adopted plate scale is 0.3\,arcsec/pxl, corresponding to a 96$^{\prime\prime} \times 72^{\prime\prime}$ field of view. The observations were obtained by chopping the signal and by nodding and jittering the telescope around the pointed position in the usual ABB$^{\prime}$A$^{\prime}$ mode. The raw data were reduced by using standard procedures for bad pixel removal and the observed field was flux calibrated by using photometric standard stars (HD29291, HD32887, HD123139). The photometric results are given in Tab.\,\ref{table:cores-FIR-assoc}, together with the IRAS/MSX results. Although these latter refer to different effective wavelengths and different epochs, ground-based values are significantly lower than IRAS/MSX determinations. These discrepancies have been remarked several times in the literature concerning YSO's (e.g. Walsh et al. 2001) and may be due to the higher environmental contamination suffered by the larger IRAS/MSX beams.

\subsection{FIR associations} \label{par:FIR-associations}
Within molecular clouds the correlation between the positions of dust emission cores and FIR point-like sources represents an important method to obtain a census of both the young stellar population and the different modalities of the star formation. To search the catalogs for sources associated to the dust cores listed in Tab.\,1 of Massi et al. (2007), a working definition of the core size has to be firstly provided. Indeed, in that Table the core size is given (column\,4), which results from the geometrical mean of the quantities $\Delta$x and $\Delta$y. These latter (see Tab.\,\ref{table:cores-NIR-assoc}, column 2) are directly provided by the adopted search algorithm (Clumpfind), and indicate the FWHM of the linear profile of the core itself, along its x (right ascension) and y (declination) axes, respectively \cite{Williams1994}. The area covered by a dust core up to the (bidimensional) FWHM flux level can thus be roughly individuated by the ellipse centered at the peak coordinates and having axes $\Delta$x and $\Delta$y. For simplicity, we will call hereinafter this ellipse as dust FWHM-ellipse (see e.g. Fig.\,\ref{Fig:MMS1}, where it is represented by the red, inner curve). Analogously, we operatively define as dust 2FWHM-ellipse that with axes 2$\Delta$x and 2$\Delta$y (red, outer ellipse in Fig.\,\ref{Fig:MMS1}).\\
Searching for catalogued sources we use this FWHM value by adopting the following criterion: an IRAS or MSX point source is considered associated to the core if its positional uncertainty ellipse overlaps (or is tangent to) the dust 2FWHM-ellipse. However, to evidence the most compelling cases, the association within one FWHM-ellipse are boldfaced in Tab.\,\ref{table:cores-FIR-assoc}. With respect to the criterion adopted in similar works (e.g. Mookerjea et al. 2004, Beltr\'an et al. 2006), for which an IRAS/MSX source is associated to a core if it lies inside 90$^{\prime \prime}$/40$^{\prime \prime}$, our criterion both takes into account the dust emission morphology and compensates for the large IRAS/MSX beam.

\subsection{NIR associations} \label{par:NIR-associations}

Both the 2MASS catalog and the IRAC2 catalog reported in Massi et al. (1999) were searched for NIR associations, i.e. the sources that dominates the cores heating. Since the NIR sources positional accuracy is by far larger than the deconvolved core size, there is no need of defining a specific criterion for the association: we simply consider all the NIR sources falling within the FWHM-ellipse of each core that present a valid flux (not an upper limit) at least in a single band ($J$, $H$, $K$). Furthermore, we have tentatively selected the most probable NIR counterpart of the dust core according to the following criteria:

\begin{enumerate}
\item Closeness to the peak of dust emission and to the FIR source possibly associated, if any.
\item Intrinsic excess in the two colours (J-H vs H-K) diagram (hereafter colour-colour diagram). This criterion makes it possible to pick up the NIR objects whose spectral energy distribution (SED) is typical of a young stellar object (YSO), and does not appear as stellar photosphere reddened by the intervening dust along the line of sight. In this context, to point out their intrinsic colour excess, we will define two loci in the colour-colour diagram (e.g. shaded regions in panel a of Fig.\,\ref{Fig:MMS1-colcol-SED}): the locus of the {\it red} objects (mainly T-Tauri), immediately to the right of the main sequence (reddened) stars and the locus of the {\it very red} sources (mainly Class I and Herbig Ae/Be protostars), to the right of the T-Tauri (reddened) stars.
\item Largest spectral index $\alpha=d\log(\lambda F_{\lambda})/d\log(\lambda)$ among those with $\alpha$\,$\geq$\,0. Sources with $\alpha <$\,0, in fact, are generally visible in the optical plates, thus this item cut off at least bright visible stars.
\end{enumerate}

\subsection{The core MMS1} \label{par:MMS1}
Here we describe a typical example of the approach adopted to detect the FIR/NIR counterparts of the $mm$-core MMS1. Similar considerations have been done for any individual $mm$-core, and all the results are provided in the Appendix A.\\
\begin{figure*}[h]
   \centering
   \includegraphics[width=10cm]{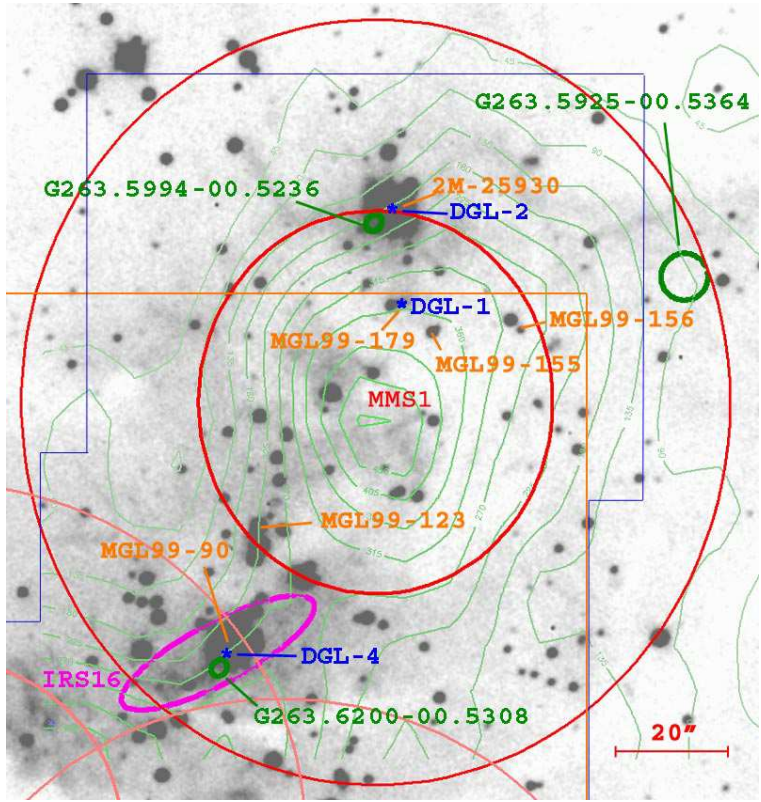}
   \caption{MMS1 field of view (center [J2000]: 08:45:32.8, -43:50:12.3). Grayscale image: $H_2$ emission; green contours: dust continuum (from 3\,$\sigma$, in steps of 3\,$\sigma$). See text for other details.}
              \label{Fig:MMS1}
\end{figure*}
As a first step, we overlay the contour map of dust emission (green contours, from 3\,$\sigma$\,=\,3\,$\times$\,15\,mJy/beam in steps of 3\,$\sigma$) and the $H_2$ narrow-band (gray-scale) image (Fig.\,\ref{Fig:MMS1}). The red ellipses centered on the $mm$-peak represent the FWHM- and the 2FWHM-ellipses within which the association with the IR sources has been searched (see Sections\,\ref{par:FIR-associations} and \ref{par:NIR-associations}). The magenta and green ellipses in Fig.\,\ref{Fig:MMS1} individuate the 3\,$\sigma$ positional uncertainties of the IRAS\footnote{Hereinafter we will adopt for the IRAS sources, when available, the shorter names IRS$\sharp$ defined in Liseau et al. (1992). Both the original names and these ones, however, are listed in Tab.\,\ref{table:cores-FIR-assoc}.} and MSX point sources, respectively, while the blue asterisks signal the position of the 10.4\,$\mu$m sources observed by Timmi2 in the field of view delimited in figure by the blue line. The 2MASS and IRAC2 NIR sources are labelled 2M $\sharp$ (following an internal numbering) and MGL99 $\sharp$ (following Massi et al. 1999), respectively. The IRAC2 field of view does not cover the whole image and is depicted as the orange rectangle.\\
Both IRS16 ($08438-4340$, corresponding to MSX G263.6200-00.5308 and DGL 4) and G263.5925-00.5364 are located outside the FWHM-ellipse and do not appear directly associated to the core, while the association is more compelling with the sources DGL 1, DGL 2 and G263.5994-00.5236.\\ 
The IRAS source coincides with a NIR young cluster and an HII region located at the center of the region bordered by the cores MMS1-2-3, almost equidistant from all of them. The NIR cluster has been already investigated in detail by Massi et al. (2003) and we complement those data with the new $N$-band observation that points out the presence of a very diffuse emission, indicated as DGL 4, corroborating the hypothesis that the FIR source can be originated by warm circumstellar matter associated with the most luminous (in the NIR) cluster member, MGL99 90.\\
Also the MSX source G263.5295-00.5364, that falls at the western border of the 2FWHM-ellipse, does not seem to be related to the dust peak. It has been detected at 8\,$\mu$m only and, presumably, such flux arises from diffuse emission, as an inspection of the MSX image suggests\footnote{It lacks of any NIR suitable counterpart (unfortunately, at that position we have no IRAC2 data) although the nearest NIR significant object, MGL99 156, 27 arcseconds apart, may contribute to the measured flux.}.\\
\begin{figure*}[h]
%   \centering
   \includegraphics[width=14cm]{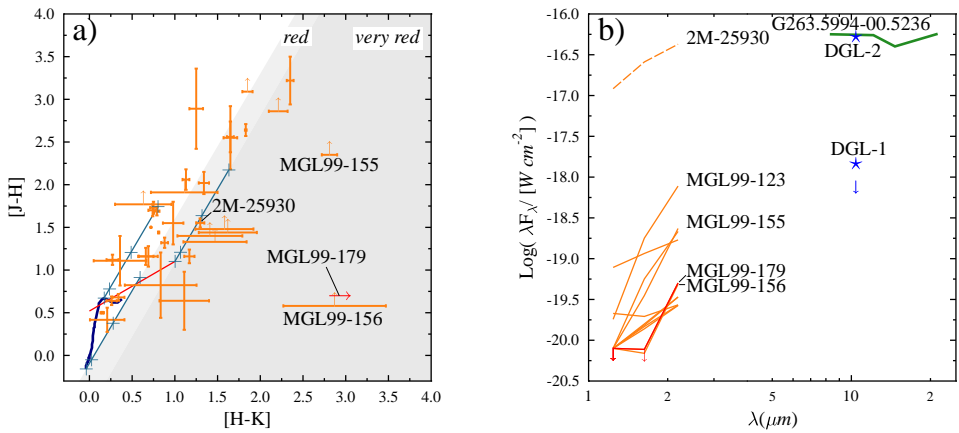}
   \caption{\textbf{a -} Colour-colour diagram of the sources falling within the FWHM-ellipse around MMS1. \textbf{b -} Spectral energy distribution of the very red sources within the FWHM-ellipse. See text for more details.}
              \label{Fig:MMS1-colcol-SED}
\end{figure*}
To feature the NIR stellar content close to the peak position we give in Fig.\,\ref{Fig:MMS1-colcol-SED}-a the colour-colour diagram of all the detected sources within the FWHM-ellipse. Here are also drawn: the locus of the main sequence, class V stars \cite{Tokunaga2000} (dark curve), the locus of the T-Tauri stars \cite{Meyer1997} (red line) and three reddening lines (blue) \cite{Rieke1985}, starting from three significant points, with four crosses indicating values of 0, 1, 5 and 10 mag of visual extinction. The separation of the sources in the two regions \textit{red} and \textit{very red} (see Sect.\,\ref{par:NIR-associations}) is also evidenced by a different shading. The arrows on the data points denote constraints on colours derived from upper limits on the NIR photometry. This diagram allows us to select the most interesting objects: in particular, the sources labelled as MGL99 156, MGL99 179 and MGL99 155 show the highest colour excess.\\
We report in Fig.\,\ref{Fig:MMS1-colcol-SED}-b the Spectral Energy Distribution (SED) of the very red stars within the FWHM-ellipse, together with the Timmi2 measured fluxes and the SED of the MSX object G263.5994-00.5236. The arrows denote again the upper limits.\\
From the SEDs, we see that MGL99 179 and MGL99 156, considering the upper limits, are the sources with the steepest spectral indexes. Moreover, (see Fig.\,\ref{Fig:MMS1}), while MGL99 179 has been identified as the counterpart of the Timmi2 source DGL 1, MGL99 156 has lack of detection in $N$, although its $K$ magnitude is comparable to that of MGL99 179.\\
This allows us to suggest that the main contributor to the millimeter flux should be the object MGL99 179\,=\,DGL 1. However, considering the richness of very red sources in the field (Tab.\,\ref{table:cores-NIR-assoc}), a contribution to the millimeter flux from multiple sources or from objects too much embedded to be investigated with the described tools, cannot be ruled out. Moreover, the HII region at the southern border of the core could provide an additional external heating by means of UV photons.
\clearpage

\section{Results} \label{par:Results}

All the results derived from the association to dust emission cores of sources from both FIR/NIR catalogs and our ground-based new observations are given in three different Tables organized as follows:

\begin{itemize}

\item[-] {\bf Table\,\ref{table:cores-FIR-assoc}} - Here all the dust cores detected in the $mm$ map (column 1) having an association with a MIR and/or FIR object are listed. The core identification follows that of the dust map \cite{Massi2007}, where the naming {\it MMS\#} refers to the resolved cores while {\it umms\#} (second part of the Table) are under-resolved (i.e. size smaller than the SIMBA beam\footnote{The circumstance of having an associated FIR source makes some of the under-resolved cores suitable targets in the next future for high spatial resolution southern facilities in the $mm$-range (e.g. ALMA).}, see Tab.\,2 of Massi et al. 2007). Then, for those cores associated with a FIR source, IRAS and MSX names, flux densities and distances from the peak ($d$) are given, along with an indicator (CC = correlation coefficient for each band, see IRAS manual) of the point-like nature of the IRAS source. For the IRAS objects is also indicated the 1.2\,$mm$ flux at the IRAS coordinates obtained by integrating the dust map within 24\,arcsec (=\,SIMBA\,HPBW) aperture. Finally, for comparison purposes with IRAS (12\,$\mu$m) and MSX (8.28 and 12.13\,$\mu$m) fluxes, the results of the Timmi2 observations (at 10.4\,$\mu$m, objects coded as DGL $\sharp$) are also reported in the last columns of this Table (again with the distances from the peak). As anticipated, the boldfaced lines correspond to FIR/MIR associations within the corresponding FWHM-ellipse, and, consequently, have to be considered as more robust cases.\\
We remark here that all the cores reported in Tab.\,\ref{table:cores-FIR-assoc} are also associated with CO clumps, with the exception of MMS29, umms1-23-24 and 25, which are outside the CO map coverage\footnote{The integrated $^{12}$CO and $^{13}$CO maps show a noticeable enhancement of the integrated emission towards umms1, although this enhancement is not fully mapped. 
%As far as those clumps that do not present any FIR association, we signal also the absence of gas clumps around MMS10-11 and umms2-3-4-5-7-10-12-13-14-22.
\label{note:umms1-CO}} \cite{Elia2007}.  

\item[-] {\bf Table\,\ref{table:cores-NIR-assoc}} - This Table lists the NIR counterparts of all the cores.
In column 2 the core size is identified through the FWHM-ellipse axes $\Delta$x and $\Delta$y (see Sect.\,\ref{par:FIR-associations}).
In column 3 the numbers of NIR red or very red sources that are located within the FWHM-ellipse are given. The census of the NIR population is based, whenever it is possible, on the IRAC2 images (Massi et al. 1999, objects coded as MGL99 $\sharp$) that are deeper than the 2MASS frames ($K$ band limit magnitude 18 instead of 14). Otherwise 2MASS images have been exploited (objects coded as 2M $\sharp$). In columns 4 to 6 the location of the most probable candidate counterpart to the $mm$ emission or to the FIR associated object is given, while the 7th column signals the morphology of the $H_2$ emission as detected in our ground-based observations.

\item[-] {\bf Table\,\ref{table:IRAS-not-assoc}} - This table lists all the IRAS sources, falling within the dust map, but not associated to any core, that present at least two valid detections (not upper limits) between 12 and 60\,$\mu$m and a flux increasing with the wavelength, namely $F_{12\mu m} < F_{25\mu m} < F_{60\mu m}$, or upper limits at 12 and 25\,$\mu$m compatible with this condition. Studying the VMR-D young population, these sources have some relevance. Indeed these red and cold sources are not randomly distributed, as the not-red sources do, but tend to be located along the gas filaments. As in Tab.\,\ref{table:IRAS-not-assoc}, we also report the 1.2\,mm flux derived at the IRAS position.
%(see Fig.\,\ref{Fig:IRAS-Fcresc-MSX}).
\end{itemize}

\clearpage

\addtolength{\voffset}{-1.5cm}
\addtolength{\headsep}{+0.9cm}

\begin{sidewaystable*}
\caption{Dust cores with associated FIR and/or MIR point sources.}
\label{table:cores-FIR-assoc}
\begin{minipage}[t][140mm]{\textwidth}
\begin{tiny}
\centering
\begin{tabular}{c | c c c c c c c c | c c c c c c | c c c c c}
\hline\hline
$mm$   &    \multicolumn{8}{c|}{IRAS associated sources $^{a}$}                                                                                               &   \multicolumn{6}{c|}{MSX associated sources$^a$}                                 &   \multicolumn{5}{c}{Timmi2 observations$^{a}$}                           \\
core     &    id                    &   $F_{12}$        &   $F_{25}$        &   $F_{60}$         &   $F_{100}$        & $F_{1.2mm}$     &  CC$^b$       &  $d$  &   id                         &   $F_{8.3}$     &   $F_{12.1}$   &   $F_{14.7}$   &   $F_{21.3}$   &   $d$   &   id   &   RA        &    Dec      &   $F_{10.4}$   &   $d$   \\
         &                          &   \multicolumn{5}{c}{(Jy)}                                                                        &               &  (")  &                              &   \multicolumn{4}{c}{(Jy)}                                         &   (")   &        &   (J2000)   &   (J2000)   &   (Jy)         &   (")   \\
\hline                                                                                                                                                                                                
MMS1     &    08438-4340            &   13.4            &   56.0            &   638.3            &   1576             &     0.203       & ECDB          &  49   &   \textbf{G263.5994-00.5236} &   \textbf{1.6}  &   \textbf{2.2}   &   \textbf{2.0}   &   \textbf{4.1}   &   29   &   \textbf{DGL 1}   &   8:45:33   &   -43:50:04   &   \textbf{0.05}   &   17   \\
         &    (IRS16)               &                   &                   &                    &                    &                 &               &       &   G263.6200-00.5308          &   1.8           &   5.1   &   6.9   &   15.6   &   51   &   \textbf{DGL 2}   &   8:45:33   &   -43:49:48   &   \textbf{1.8}   &   33   \\
         &                          &                   &                   &                    &                    &                 &               &       &   G263.5925-00.5364          &   0.5           &   $<$0.7   &   $<$0.5   &   $<$1.6   &   56   &   DGL 4   &   8:45:36   &   -43:51:02   &   diffuse   &   $\sim$49   \\
MMS2     &                          &                   &                   &                    &                    &                 &               &       &   \textbf{G263.6338-00.5497} &   0.4           &   \textbf{$<$6.9}   &   \textbf{$<$5.4}   &   \textbf{1.6}   &   25   &   \textbf{DGL 3}   &   8:45:34   &   -43:52:26   &   \textbf{0.25}    &   26   \\
MMS3     &    08438-4340            &   13.4            &   56.0            &   638.3            &   1576             &     0.203       & ECDB          &  55   &   \textbf{G263.6385-00.5217} &   \textbf{0.9}  &   \textbf{1.3}   &   \textbf{0.8}   &   \textbf{1.4}   &   20   &   \textbf{DGL 5}   &   8:45:38   &   -43:51:14   &   \textbf{0.08}   &   26   \\
         &    (IRS16)               &                   &                   &                    &                    &                 &               &       &   G263.6366-00.5148          &   0.3           &   $<$0.9   &   $<$0.7   &   $<$1.9   &   39   &   DGL 4   &   8:45:36   &   -43:51:02   &   diffuse   &   $\sim$55   \\
         &                          &                   &                   &                    &                    &                 &               &       &   G263.6329-00.5127          &   0.4           &   0.7   &   $<$0.9   &   $<$2.7   &   46   &      &      &      &      &      \\
         &                          &                   &                   &                    &                    &                 &               &       &   G263.6200-00.5308          &   1.8           &   5.1   &   6.9   &   15.6   &   54   &      &      &      &      &      \\
MMS4$^c$ &    \textbf{08448-4343}   &   \textbf{8.7}    &   \textbf{88.1}   &   \textbf{326.6}   &   \textbf{1005}    & \textbf{0.984}  & \textbf{AABB} &  7    &   \textbf{G263.7759-00.4281} &   \textbf{7.5}  &   \textbf{10.1}   &   \textbf{13.8}   &   \textbf{65.5}   &   5   &   \textbf{MGL99 57}$^d$   &   8:46:35   &   -43:54:31   &   \textbf{8.79}   &   7   \\
         &    \textbf{(IRS17)}      &                   &                   &                    &                    &                 &               &       &   G263.7733-00.4151          &   1.2           &   1.3   &   0.5   &   2.6   &   53   &   \textbf{MGL99 25}$^d$   &   8:46:34   &   -43:54:50   &   \textbf{0.21}   &   15   \\
         &                          &                   &                   &                    &                    &                 &               &       &   G263.7867-00.4437          &   0.3           &   0.8   &   $<$0.4   &   $<$1.2   &   63   &   \textbf{MGL99 40}$^d$   &   8:46:33   &   -43:54:39   &   \textbf{0.14}   &   17   \\
MMS12    &    \textbf{08470-4321}   &   \textbf{44.9}   &   \textbf{130.1}  &   \textbf{342.6}   &   \textbf{406.9}   & \textbf{0.331}  & \textbf{AAAA} &  11   &   \textbf{G263.7434+00.1161} &   \textbf{30.8} &   \textbf{52.6}   &   \textbf{70.3}   &   \textbf{93.5}   &   3   &   \textbf{DGL 7}   &   8:48:49   &   -43:32:29   &   \textbf{18.3}   &   5   \\
         &    \textbf{(IRS19)}      &                   &                   &                    &                    &                 &               &       &                              &                 &      &      &      &      &      &      &      &      &      \\
MMS18    &    08472-4326A           &   0.9             &   0.9             &   10.6             &   $<$406.9         &  0.044          & BABF          &  28   &   \textbf{G263.8432+00.0945} &   \textbf{0.3}  &   \textbf{$<$0.7}   &   \textbf{$<$0.5}   &   \textbf{$<$1.3}   &   16   &   \textbf{DGL 8}   &   8:49:03   &   -43:37:55   &   \textbf{0.07}   &   19   \\
MMS20    &    08474-4323            &   1.5             &   1.1             &   $<$106.4         &   125.9            &  $<$0.005       & BCAD          &  39   &   G263.8221+00.1494          &   0.1           &   $<$0.9   &   $<$0.6   &   $<$1.9   &   37   &   \textbf{DGL 9}   &   8:49:12   &   -43:35:52   &   \textbf{0.08}   &   28   \\
MMS21    &    \textbf{08474-4325}   &   \textbf{$<$0.3} &   \textbf{1.0}    &   \textbf{$<$16.4} &   \textbf{57.5}    & \textbf{0.207}  & \textbf{DAHD} &  7    &                              &                 &      &      &      &      &   DGL 9   &   8:49:12   &   -43:35:52   &   0.08    &   39   \\
MMS22    &    \textbf{08476-4306}   &   \textbf{5.7}    &   \textbf{44.0}   &   \textbf{216.3}   &   \textbf{503.7}   & \textbf{0.337}  & \textbf{AAAB} &  10   &   \textbf{G263.6177+00.3652} &   \textbf{3.9}  &   \textbf{6.0}   &   \textbf{7.6}   &   \textbf{27.7}   &   12   &   \textbf{DGL 11}   &   8:49:26   &   -43:17:12   &   \textbf{2.01}   &   12   \\
         &    \textbf{(IRS20)}      &                   &                   &                    &                    &                 &               &       &                              &                 &      &      &      &      &   DGL 10   &   8:49:26   &   -43:17:21   &   0.02   &   19   \\
MMS24    &    08476-4306            &   5.7             &   44.0            &   216.3            &   503.7            &  0.337          & AAAB          &  48   &   G263.6280+00.3847          &   0.3           &   $<$0.6   &   $<$0.5   &   $<$1.3   &   28   &      &      &      &   $<$0.03   &      \\
         &    (IRS20)               &                   &                   &                    &                    &                 &               &       &                              &                 &      &      &      &      &      &      &      &      &      \\
MMS25    &    \textbf{08477-4359}   &   \textbf{9.0}    &   \textbf{26.3}   &   \textbf{317.0}   &   \textbf{580.8}   & \textbf{0.174}  & \textbf{BAAA} &  20   &   \textbf{G264.3225-00.1857} &   \textbf{4.7}  &   \textbf{5.0}   &   \textbf{2.3}   &   \textbf{8.0}   &   19   &   \textbf{DGL 12}   &   8:49:33   &   -44:10:60   &   \textbf{0.21}   &   30   \\
         &    \textbf{(IRS21)}      &                   &                   &                    &                    &                 &               &       &                              &                 &      &      &      &      &      &      &      &      &      \\
MMS26    &    \textbf{08477-4359}   &   \textbf{9.0}    &   \textbf{26.3}   &   \textbf{317.0}   &   \textbf{580.8}   & \textbf{0.174}  & \textbf{BAAA} &  16   &   \textbf{G264.3225-00.1857} &   \textbf{4.7}  &   \textbf{5.0}   &   \textbf{2.3}   &   \textbf{8.0}   &   16   &   \textbf{DGL 12}   &   8:49:33   &   -44:10:60   &   \textbf{0.21}   &   6   \\
         &    \textbf{(IRS21)}      &                   &                   &                    &                    &                 &               &       &                              &                 &      &      &      &      &      &      &      &      &      \\
MMS27    &                          &                   &                   &                    &                    &                 &               &       &                              &                 &      &      &      &      &   \textbf{DGL 13}   &   8:49:36   &   -44:11:46   &   \textbf{0.03}   &   16   \\
MMS28    &    \textbf{08483-4305}   &   \textbf{1.5}    &   \textbf{2.5}    &   \textbf{$<$39.0} &   \textbf{189.2}   & \textbf{0.049}  & \textbf{EEFD} &  14   &   \textbf{G263.6909+00.4713} &   \textbf{0.1}  &   \textbf{0.6}   &   \textbf{$<$0.7}   &   \textbf{$<$2.0}   &   11   &   not obs.   &      &      &      &      \\
MMS29    &    08483-4305            &   1.5             &   2.5             &   $<$39.0          &   189.2            & 0.049           & EEFD          &  45   &                              &                 &      &      &      &      &   not obs.   &      &      &      &      \\
\hline                                                                                                                                                          
umms1    &    \textbf{08446-4331}   &   \textbf{$<$1.1} &   \textbf{0.5}    &   \textbf{7.4}     &   \textbf{42.8}    & \textbf{0.160}  & \textbf{GBBB} &  2    &                              &                 &      &      &      &      &   not obs.   &      &      &      &      \\
umms8    &    08458-4332            &   1.1             &   2.7             &   17.8             &   52.7             & 0.017           & CABA          &  24   &                              &                 &      &      &      &      &      &      &      &   $<$0.03   &      \\
umms9    &    \textbf{08458-4332}   &   \textbf{1.1}    &   \textbf{2.7}    &   \textbf{17.8}    &   \textbf{52.7}    & \textbf{0.017}  & \textbf{CABA} &  29   &                              &                 &      &      &      &      &      &      &      &   $<$0.03   &      \\
umms11   &                          &                   &                   &                    &                    &                 &               &       &   \textbf{G263.7651-00.1572} &   \textbf{0.2}  &   \textbf{$<$0.5}   &   \textbf{0.5}   &   \textbf{$<$1.2}   &   11   &   \textbf{DGL 6}   &   8:47:43   &   -43:43:48   &   \textbf{0.07}   &   10   \\
umms16   &    \textbf{08464-4335}   &   \textbf{$<$0.3} &   \textbf{$<$0.3} &   \textbf{6.6}     &   \textbf{$<$47.2} & \textbf{0.160}  & \textbf{-JB-} &  19   &                              &                 &      &      &      &      &      &      &      &   $<$0.03   &      \\
umms23   &                          &                   &                   &                    &                    &                 &               &       &   \textbf{G263.5672+00.4036} &   \textbf{0.1}  &   \textbf{$<$0.9}   &   \textbf{$<$0.7}   &   \textbf{$<$1.9}   &   17   &   not obs.   &      &      &      &      \\
umms24   &                          &                   &                   &                    &                    &                 &               &       &   \textbf{G263.5622+00.4185} &   \textbf{0.2}  &   \textbf{$<$0.7}   &   \textbf{$<$0.5}   &   \textbf{$<$1.5}   &   13   &   not obs.   &      &      &      &      \\
umms25   &                          &                   &                   &                    &                    &                 &               &       &   G263.5622+00.4185          &   0.2           &   $<$0.7   &   $<$0.5   &   $<$1.5   &   19   &   not obs.   &      &      &      &      \\

\hline \\
\end{tabular}

\end{tiny}
\begin{tiny}
Notes to the table: the $mm$ sources labeled as umms$\sharp$ are not resolved by the SIMBA HPBW (see text and Massi et al. 2007).\\
$^a$Bold faced sources are more compelling associations (within the FWHM-ellipse of the core, see text).\\
$^b$Point source correlation coefficient encoded as alphabetic characters (A=100$\%$, B=99$\%$, ..., N=87$\%$) according to the IRAS Catalogs and Atlases Explanatory Supplement \cite{IRAS-PSC1988}.\\
$^c$This core has been divided into two components by Giannini et al. (2005).\\
$^d$Names following the numbering convention used in Massi et al. (1999). A description of the Timmi2 observations of these objects is given in Giannini et al. (2005).\\
\end{tiny}
\end{minipage}
\end{sidewaystable*}

\begin{table*}
\caption{NIR sources and $H_2$ emission associated with dust cores.}
\label{table:cores-NIR-assoc}
\begin{minipage}[130mm]{\textwidth}
\begin{tiny}
\centering
\begin{tabular}{c | c | c | c c c | c }
\hline\hline
$mm$ core  & FWHM                     & $\sharp$ of NIR$^a$                          &  \multicolumn{3}{c|}{Counterparts candidates}  &  $H_2$ emission  \\
name       & $\Delta$x-$\Delta$y    & \textit{red} / \textit{very red} sources$^b$   &    name   & RA              &  Dec             &  morphology$^c$      \\
           & (")                      & within the FWHM-ellipse                      &           &    (J2000)      &     (J2000)      &                  \\
\hline
MMS1$^d$   & 59-64                    & 5 / 8 $^e$                                   & MGL99 179     & 08:45:32.86     & -43:50:03.80     &  \textbf{diffuse}, knots          \\
MMS2$^d$   & 59-55                    & 1 / 3 $^e$                                   & MGL99 25      & 08:45:35.93     & -43:51:45.60     &  \textbf{diffuse, jet-like}       \\
MMS3$^d$   & 61-58                    &  0 / 1 $^e$                                  & MGL99 36      & 08:45:39.21     & -43:51:34.70     &  \textbf{diffuse}                 \\
MMS4$^d$   & 82-71                    & 19 / 43 $^f$                                 & MGL99 57      & 08:46:34.77     & -43:54:30.63     & \textbf{diffuse, jet-like, knots} \\
MMS5       & 40-31                    &  0 / 0                                       &           &                 &                  &  \textbf{diffuse}                 \\
MMS6       & 27-23                    &  0 / 0                                       &           &                 &                  &  diffuse                          \\
MMS7       & 29-37                    &  1 / 4                                       &           &                 &                  &  -                                \\
MMS8       & 24-26                    &  0 / 1                                       & 2M 9671   & 08:48:39.13     & -43:31:31.36     &  -                                \\
MMS9       & 32-36                    &  0 / 0                                       &           &                 &                  &  -                                \\
MMS10      & 25-28                    &  0 / 0                                       &           &                 &                  &  -                                \\
MMS11      & 36-24                    &  0 / 1                                       & 2M 14732  & 08:48:46.54     & -43:37:44.02     &  -                                \\
MMS12$^d$  & 57-37                    &  8 / 20 $^f$                                 & MGL99 49      & 08:48:48.51     & -43:32:29.08     &  \textbf{diffuse, knots}          \\ 
MMS13      & 41-20                    & 2 / 0 $^f$                                   & MGL99 2       & 08:48:50.04     & -43:33:19.47     &  -                                \\
MMS14      & 29-38                    & 0 / 0                                        &           &                 &                  &  -                                \\
MMS15      & 33-20                    & 0 / 0                                        &           &                 &                  &  -                                \\
MMS16      & 23-36                    & 0 / 0                                        &           &                 &                  &  \textbf{diffuse, jet-like}       \\
MMS17      & 67-37                    & 0 / 2                                        &           &                 &                  &  \textbf{jet-like}                \\
MMS18$^d$  & 36-36                    & 1 / 0                                        &           &                 &                  &  \textbf{diffuse}                 \\
MMS19      & 26-60                    & 3 / 2                                        & 2M 36076  & 08:49:08.49     & -43:35:37.79     &  -                                \\
MMS20$^d$  & 38-32                    & 0 / 0                                        &           &                 &                  &  knots                            \\
MMS21$^d$  & 54-57                    & 0 / 3                                        & 2M 29953  & 08:49:13.39     & -43:36:29.20     &  knots                            \\
MMS22$^d$  & 35-37                    & 9 / 13 $^f$                                  & MGL99 98      & 08:49:26.23     & -43:17:11.11     &  \textbf{diffuse, jet-like, knots}\\
MMS23      & 39-22                    & 0 / 0                                        &           &                 &                  &  \textbf{diffuse}                 \\
MMS24$^d$  & 35-25                    & 0 / 3 $^f$                                   & MGL99 90      & 08:49:32.27     & -43:17:14.43     &  -                                \\
MMS25$^d$  & 52-86                    & 3 / 5                                        &           &                 &                  &  \textbf{diffuse}                 \\
MMS26$^d$  & 60-67                    & 1 / 8                                        &           &                 &                  &  \textbf{diffuse}                 \\
MMS27$^d$  & 67-48                    & 1 / 2                                        &           &                 &                  &  \textbf{knots}                   \\
MMS28$^d$  & 30-31                    & 0 / 0                                        &           &                 &                  &  -                                \\
MMS29$^d$  & 40-36                    & 1 / 1                                        & 2M 36339  & 08:50:11.04     & -43:17:10.74     &  \textbf{knots}                   \\
\hline                                                                                                                                                                    
                                                                                                                                                                          
umms1$^d$  & $<$24-24                 & 0 / 0                                        &           &                 &                  &  not observed                     \\
umms2      & $<$24-24                 & 0 / 0                                        &           &                 &                  &  -                                \\
umms3      & $<$24-24                 & 0 / 0                                        &           &                 &                  &  -                                \\
umms4      & $<$24-24                 & 0 / 1                                        &           &                 &                  &  -                                \\
umms5      & $<$24-24                 & 1 / 0                                        &           &                 &                  &  -                                \\
umms6      & $<$24-24                 & 1 / 0                                        &           &                 &                  &  -                                \\
umms7      & $<$24-24                 & 0 / 0                                        &           &                 &                  &  -                                \\
umms8$^d$  & $<$24-24                 & 1 / 1                                        & 2M 16128  & 08:47:37.87     & -43:43:42.41     &  -                                \\
umms9$^d$  & $<$24-24                 & 0 / 0                                        &           &                 &                  &  -                                \\
umms10     & $<$24-24                 & 0 / 0                                        &           &                 &                  &  -                                \\
umms11$^d$ & $<$24-24                 & 0 / 2                                        & 2M 9173   & 08:47:42.93     & -43:43:48.05     &  -                                \\
umms12     & $<$24-24                 & 0 / 0                                        &           &                 &                  &  -                                \\
umms13     & $<$24-24                 & 0 / 0                                        &           &                 &                  &  -                                \\
umms14     & $<$24-24                 & 0 / 0                                        &           &                 &                  &  -                                \\
umms15     & $<$24-24                 & 0 / 0                                        &           &                 &                  &  -                                \\
umms16$^d$ & $<$24-24                 & 0 / 0                                        &           &                 &                  &  \textbf{jet-like}                \\
umms17     & $<$24-24                 & 0 / 0                                        &           &                 &                  &  jet-like                         \\
umms18     & $<$24-24                 & 0 / 0                                        &           &                 &                  &  \textbf{jet-like}                \\
umms19     & $<$24-24                 & 1 / 0                                        & 2M 10742  & 08:48:33.95     & -43:30:47.20     &  \textbf{knot}                    \\
umms20     & $<$24-24                 & 0 / 0                                        &           &                 &                  &  knot                             \\
umms21     & $<$24-24                 & 0 / 3                                        & 2M 16489  & 08:48:37.03     & -43:13:53.63     &  -                                \\
umms22     & $<$24-24                 & 0 / 0                                        &           &                 &                  &  -                                \\
umms23$^d$ & $<$24-24                 & 0 / 1                                        & 2M 11799  & 08:49:24.56     & -43:13:15.49     &  -                                \\
umms24$^d$ & $<$24-24                 & 0 / 0                                        &           &                 &                  &  -                                \\
umms25$^d$ & $<$24-24                 & 0 / 0                                        &           &                 &                  &  -                                \\
umms26     & $<$24-24                 & 0 / 0                                        &           &                 &                  &  -                                \\
\hline
\end{tabular}
\end{tiny}

\begin{tiny}
$^a$Where not specified, the reported numbers refer to 2MASS data.\\
$^b$The terms \textit{red} and \textit{very red} refer to different regions of the colour-colour diagram (see text and e.g. Fig.\,\ref{Fig:MMS1-colcol-SED}-a). \\
$^c$Bold-faced if inside the FWHM-ellipse.\\
$^d$Cores presenting a MIR or FIR association (see Tab.\,\ref{table:cores-FIR-assoc}).\\
$^e$Core only partially covered by IRAC2 observations.\\
$^f$Core fully covered by IRAC2 observations.\\
\end{tiny}
\end{minipage}
\end{table*}

\clearpage

\addtolength{\voffset}{+1.5cm}
\addtolength{\headsep}{-0.5cm}

\begin{table*}
\caption{IRAS point sources with increasing fluxes not associated with any core.}
\label{table:IRAS-not-assoc}
\begin{minipage}[130mm]{\textwidth}
%\centering
\begin{tabular}{c c c c c c c}
\hline\hline
          id   & $F_{12}$  & $F_{25}$& $F_{60}$& $F_{100}$ & $F_{1.2mm}$ & CC$^a$	\\
          	   &	\multicolumn{5}{c}{(Jy)}                             &          \\
  \hline
  08440-4253   &  $<$0.3   &    0.2  &  1.2    & $<$30.5   & $<$0.005    & -BBD     \\
  08461-4314   &  0.7      &    0.8  &  5.9    & $<$48.5   & 0.003       & BCDB     \\
  08475-4255   &  $<$0.4   &    0.4  &  3.3    & 45.6      & 0.020       & EDDB     \\
  08475-4311   &  $<$0.5   &    0.5  &  9.4    & $<$37.8   & 0.020       & DEC-     \\
  08481-4258   &  $<$0.4   &    1.1  &  7.6    & $<$34.8   & 0.008       & BDC-     \\
  08459-4338   &  $<$0.2   &    0.6  &  6.5    & 38.3      & 0.005       & -DCB     \\
  08442-4328   &  $<$1.4   &    2.2  &  41.3   & 113.9     & 0.020       & HDCA     \\
  08448-4341   &  1.3      &    6.6  &  $<$327 & $<$1005   & $<$0.005    & DAC-     \\
  08468-4330   &  $<$0.3   &    0.4  &  3.4    & $<$31.6   & 0.010       & FDD-     \\
  08491-4310   &  0.5      &    0.6  &  7.6    & 38.6      & $<$0.005    & EBCC     \\
  08496-4320   &  0.6      &    0.7  &  7.6    & $<$37.1   & $<$0.005    & BCBH     \\
  08463-4343   &  $<$0.3   &    0.5  &  7.4    & $<$43.5   & $<$0.005    & NBDG     \\
  08478-4403   &  0.5      &    0.5  &  $<$5.4 & $<$580.8  & $<$0.005    & CCJG     \\

\hline
\end{tabular}
\end{minipage}
\\

\begin{tiny}
$^a$See note $b$ in Tab.\,\ref{table:cores-FIR-assoc}.\\
\end{tiny}

\end{table*}

\section{Discussion} \label{par:Discussion}

\subsection{IR counterparts of dust cores} \label{par:Discussion-IR-counterpart-of-dust-cores}
The starting information is the sample of 29 well resolved and 26 unresolved dust cores \cite{Massi2007}. As illustrated in the Appendix\,A, some of them are not isolated, but belong to more complex structures of dust emission that present different cores (e.g. the one composed by MMS7, umms13-14-15, Fig.\,\ref{Fig:MMS7-umms13-14-15-colcol}); other are elongated structures formed by aligned {\it knots} (e.g. both MMS5, MMS6, umms6 of Fig.\,\ref{Fig:MMS5-6-umms6} and umms23-24-25 of Fig.\ref{Fig:umms23-24-25}). These multiple core structures often present associated FIR counterpart(s) that tend(s)  to be located in a position intermediate between the individual cores. This is the reason why in Tab.\,\ref{table:cores-FIR-assoc} the same IRAS/MSX source is sometimes assigned to two different cores. Conversely, the NIR counterparts distribution is more clearly defined because of the increased spatial resolution at such wavelengths. In Appendix A all the observational details pertaining to each individual dust core are presented; however we can draw here some general remarks. Firstly, all the information given in Tabs.\,\ref{table:cores-FIR-assoc} and \ref{table:cores-NIR-assoc} are statistically summarized in Tab.\,\ref{table:peaks-popul}, where, among the 29 resolved cores (MMS) only 8 are clearly associable to a MIR/FIR source (with or without a near-IR counterpart); 7 have some MIR/FIR source in their neighbourhood, but additional evidences (i.e. the presence of $H_2$ jets) point back to an embedded object not (yet) detected; 7 additional cores seem to be associated with very red NIR objects; the 7 remaining cores do not present any sign of star formation activity at the current instrumental sensitivity. If the same approach is applied to the sample of the 26 unresolved cores (umms), we obtain that 4 are associable to a MIR/FIR source, 3 to a NIR counterpart, 3 to jet-like structures, and 16 appear as inactive sites. Substantially, both MMS and umms present a similar statistics of associated categories, although the latter sample is more widely dominated by objects that could be artifacts, caused by the searching algorithm, or sites harbouring  weak IR counterparts. The total number of resolved cores associable to an IR object (irrespective of being NIR or FIR sources) is 15 (column 6 of Tab.\,\ref{table:peaks-popul}) with respect to the 14 unassociated cores (column 7): such percentages are in full agreement with those found in other galactic surveys of star forming regions (e.g. Yonekura et al. 2005,  Mookerjea et al. 2004,  Beltr\'an et al. 2006). \\
Such categorization, however, does not necessarily reflect a property intrinsic to the cores themselves, but is likely the product of the limiting instrumental sensitivities of the considered facilities. In fact, recent results of SPITZER MIR surveys have substantially modified the percentage of active vs inactive cores in favour of the first ones (e.g. Young et al. 2004). Such a caveat should be taken into account when drawing general conclusions from our analysis. 

%%%%   TABLE  4 %%%%%%%%%%%%%%%%%%%%%%%%%%%%%%%%%%%
\begin{table*}
\caption[]{Statistics about the dust cores population. 
    \label{table:peaks-popul}}
\begin{center}
\begin{tabular}{c|ccccccc}
\hline\hline
%\hline \\[-5pt]
Cores & \multicolumn{7}{c}{Associated with} \\
\hline
 & MIR/FIR & only very red &  $H_2$ jet  & no source & NIR or     &  no source or & embedded \\
 &         & NIR object    &             &           & FIR object &  $H_2$ only   & cluster  \\
\cline{2-8}
29 resolved     &  8   &  7  &  7  &  7    &  15 & 14 & 6 \\
26 unresolved &  4   &  3  &  3  &  16  & 7 & 19 &  2\\
\hline %[-5pt]

\end{tabular}
\end{center}
\end{table*}
%%%%%%%%%%%%%%%%%%%%%%%%%%%%%%%%%%%%%%%%%%%%%%

\subsection{Star formation modalities and evolutionary stages}

\begin{figure*}[h]
   \centering
\includegraphics[width=12cm]{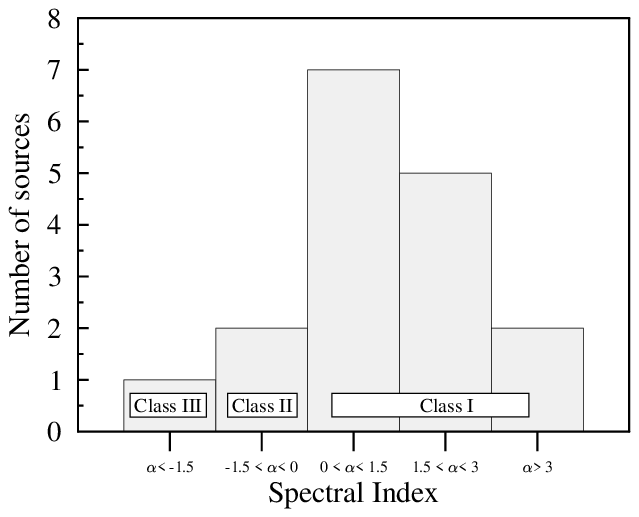}
   \caption{Spectral index ($\alpha_{2-10}$) distribution of core counterparts. In cases of multiple estimates of the 10\,$\mu$m flux (Timmi2, MSX, IRAS) the value obtained with better spatial resolution has been used to determine $\alpha$.}
   \label{Fig:N-vs-SpInd}%
\end{figure*}

Different modalities of star formation are simultaneously active in VMR-D. Such a co-existence is confirmed by the presence of 8 clusters (see last column of Tab.\,\ref{table:peaks-popul}) and by the remaining cases of isolated star formation. This twofold modality, already recognized in Orion (e.g. Chen \& Tokunaga 1994) and now in VMR-D as well, seems to be a feature of all the regions where intermediate and high-mass stars form. Indeed, it is likely that the lacking detection of the isolated mode in far and massive star forming regions is only due to limitations on sensitivity and spatial resolution.\\ 
To investigate whether or not the different cores of VMR-D harbour protostellar objects in different evolutionary stages we have constructed the distribution of the spectral slope of the sources associated to the cores (see Fig.\,\ref{Fig:N-vs-SpInd}). For each source the slope $\alpha$ is calculated through the relationship $\alpha=\Delta \log(\lambda_i F_{\lambda_i})/\Delta \log(\lambda_i)$, between the wavelengths $\lambda_ 1$ and $\lambda_ 2$, corresponding to about 2 and 10\,$\mu$m, respectively. A certain degree of inhomogeneity is introduced by the fact that the flux attributed to the 10$\mu$m band corresponds to that detected by different instruments (MSX, Timmi2, IRAS) operating at different effective wavelengths (8.28, 10.4 and 12\,$\mu$m, respectively). These differences affect only the details of the slope distribution, but do not alter its significance, as proved by the five sources having a multiple detection. Fig.\,\ref{Fig:N-vs-SpInd} illustrates how the sample of the sources associated to the dust cores presents a distribution highly peaked at values $0 < \alpha < 3$, typical of Class\,I sources. It is worthwhile noting that the NIR contribution to the slope generally relies on 2MASS data, that provide a $K$ band limiting magnitude of about 14\,mag; deeper NIR surveys (see e.g. Giannini et al. 2005), could make it possible to find weaker NIR counterparts, increasing the number of sources with high $\alpha$ values. As a result, the presented bar graph (Fig.\ref{Fig:N-vs-SpInd}) could be shifted towards larger $\alpha$ values. As expected, the sources associated to the dust cores are essentially Class\,I objects, although the distribution presents a significant tail toward the less evolved objects, whose slope is greater than 3.\\
In Tab.\,\ref{table:IRAS-not-assoc} we also  list all the IRAS sources lying inside the region mapped in the 1.2\,mm continuum and showing a flux increasing with wavelength but not directly associated to any dust core (see Sect.\,\ref{par:Results}). These sources, with only few exceptions, tend to be distributed all along both the diffuse emission detected by MSX at 8.3\,$\mu$m and the gas filaments.\\
\begin{figure*}[h]
   \centering
\includegraphics[width=14cm]{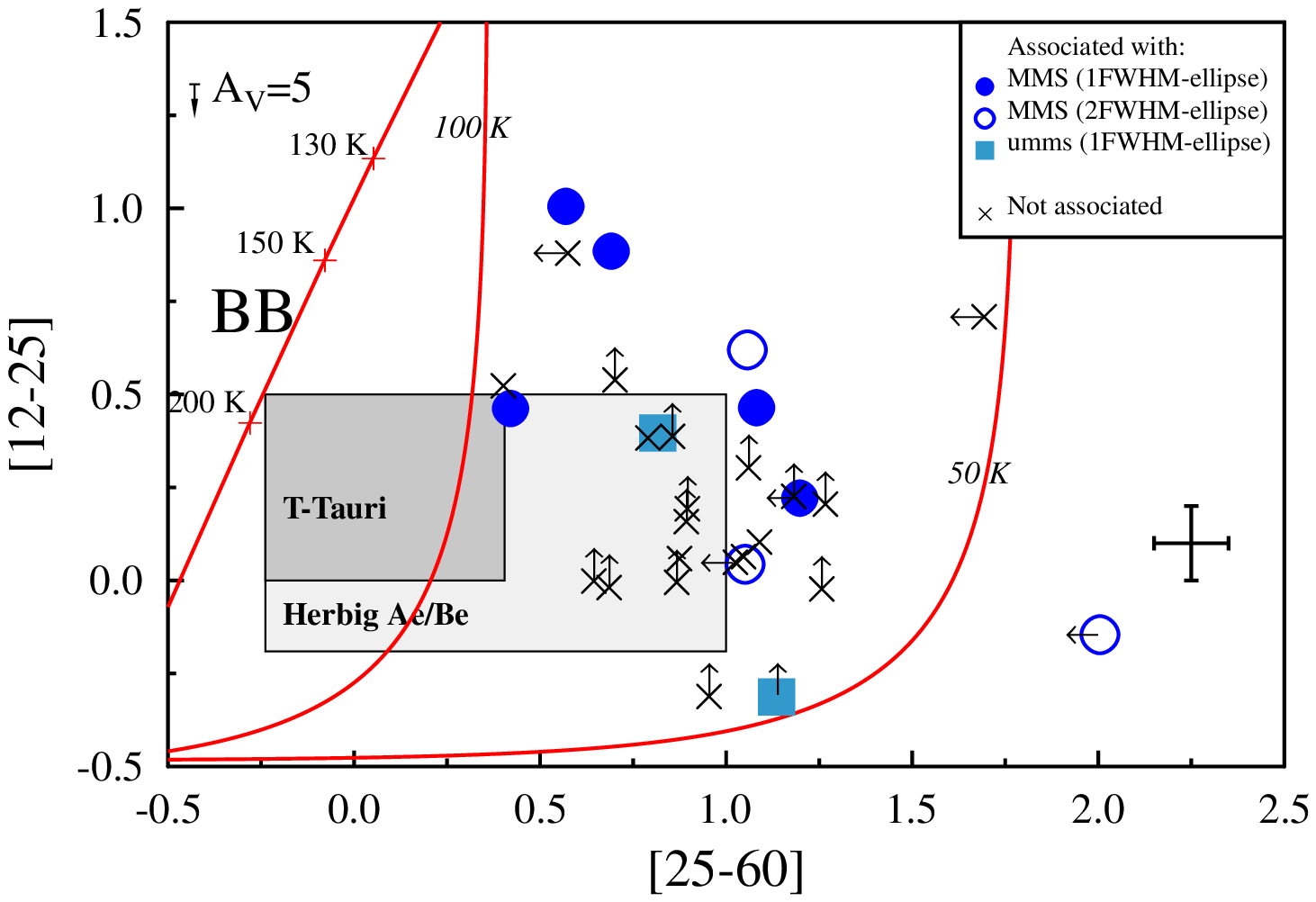}
   \caption{Two colours diagram for all the IRAS sources listed in Tabs.\,\ref{table:cores-FIR-assoc} and \ref{table:IRAS-not-assoc}. The mean error bars, the extinction vector corresponding to 5\,mag of visual extinction and the locus of blackbodies (leftmost line) are indicated (see text for more details).}
   \label{Fig:IRAS-colcol}
\end{figure*}
\begin{figure*}[h]
   \centering
\includegraphics[width=14cm]{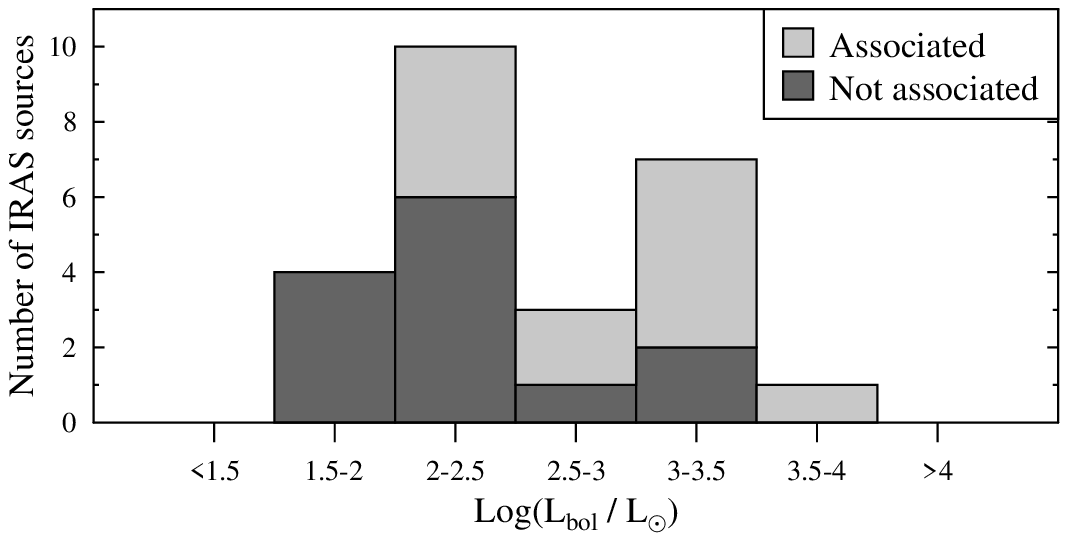}
   \caption{Luminosity distribution of both associated and not associated IRAS sources (Tabs.\,\ref{table:cores-FIR-assoc} and \ref{table:IRAS-not-assoc}).}
   \label{Fig:Istogr-assoc-vs-nonassoc}
\end{figure*}
To evaluate the intrinsic nature of all the  IRAS sources discussed so far (both associated or not), we present, in Fig.\,\ref{Fig:IRAS-colcol}, an IRAS two colours diagram, [12-25] vs [25-60], with all those sources. As expected, the objects not associated to the cores have a greater number of upper limits (especially at 12\,$\mu$m), but tend to occupy the same region of the plot where the associated sources are located. This common region pertains to sources that can be described as a two dust components system, one at 1000\,K and the other in between 50 and 100\,K, with variable relative contributions. The IRAS selected sources have colours definitely redder and colder than those pertaining to pre-main sequence stars (T-Tauri and Herbig Ae/Be; see, e.g., Berrilli et al. 1992), apart from few cases. While the absence of IRAS sources with the colours of T-Tauri stars in our sample is probably due to an observational bias (in fact the IRAS detectability limit is $>0.5\,L_{\sun}$ in Taurus, corresponding to $> 10\,L_{\sun}$ in VMR-D), the doubtful presence of Herbig Ae/Be appears to be an intrinsic property of VMR-D. Indeed, Herbig Ae/Be located inside dust cores having masses comparable to those belonging to the VMR-D ones have been already found at distances of 1\,Kpc or less (e.g. Henning et al. 1998). This difference likely reflects different star formation histories, being in VMR-D a shorter time ($\approx 10^{6}\,$yr) elapsed from the earliest collapse events \cite{Massi2000} with respect to other studied massive clouds.\\
We have calculated the bolometric luminosity, $L_{bol}$, for all the IRAS sources from 12$\mu$m to 1.2\,mm. The bar graph in Fig.\,\ref{Fig:Istogr-assoc-vs-nonassoc} shows our result: the sources associated to the dust cores (both resolved and not) are, on average, objects of intermediate luminosity ($L_{bol} \sim 10^3 L_{\odot}$), while the unassociated FIR sources, even showing similar SEDs, tend to be lower luminosity objects ($L_{bol} \sim 10^2 L_{\odot}$). Therefore, we can firmly conclude that massive star formation ($L_{bol} > 10^4 L_{\sun}$) does not occur in VMR-D, and that our sample of Class I sources is not contaminated by ultracompact HII regions, which would be indistinguishable based on their FIR colours alone \cite{Wood1989}.\\
An attempt to search for Class\,0 objects within our sample of IRAS sources has also been done by applying the criterion proposed by Andr\'e et al. (1993) for low-mass protostars: $L_{bol}/L_{1.3mm} \lesssim 2 \times 10^4$. None of the selected IRAS sources (Tabs.\,\ref{table:cores-FIR-assoc} and \ref{table:IRAS-not-assoc}) strictly satisfies this criterion\footnote{Hatchell et al. (2007) reduces such limiting ratio to $L_{bol}/L_{1.3mm} \lesssim 3 \times 10^3$ and underlines how this indicator should vary with envelope mass.}. It should be said, however, that overestimates of the IRAS fluxes result in overestimates of the $L_{bol}/L_{1.3mm}$ ratio. Thus, leaving any quantitative approach, we point out that three sources of our sample (08446-4331, 08474-4325 and 08464-4335) show a ratio $L_{bol}/L_{1.3mm}$ one order of magnitude less than the others and thus they are likely the youngest objects in the field. This hypothesis is strongly supported by {\it (i)} the tight association of these sources with the $mm$ cores umms1, MMS21 and umms16, respectively; {\it (ii)} by the lack of a measured flux at 12$\mu$m\footnote{For the core MMS21 with the associated source 08474-4325 we report in Tab.\,\ref{table:cores-NIR-assoc} a candidate NIR counterpart, but such association is quite questionable (see Appendix A and Fig.\,\ref{Fig:MMS21}).}, and, but only for the last case, {\it (iii)} by the presence of $H_2$ jet-like emission crossing the dust peak (no $H_2$ images are available for the core umms1).

\subsection{$H_2$ survey }

A powerful approach to indirectly probe ongoing star formation activity is to investigate over the presence and characteristics of collimated flows, which commonly extend far from the most  embedded regions where the young protostars are located. For this reason, we have complemented our search for infrared counterparts of $mm$ cores with a survey in the narrow-band filter centered at the $H_2$\,2.12\,$\mu$m line, which represents the main NIR cooling channel of shock-excited gas at thousands of Kelvin (e.g. Gredel 1994). Our observations have been also aimed to define the occurrence of the jet phenomenon in intermediate-mass star forming regions and to understand whether jets in clusters are associated with the massive cores or with some closeby low-mass component. A detailed description of the $H_2$ emission detected in the various fields is given in Appendix A, while here we briefly comment on some general aspects. Out of 54 investigated fields, we find a positive detection (in the form of jet-like morphologies, knots and diffuse emission) in 23 fields, the large majority (18) being related to resolved cores.  We note that the occurrence of jets in proximity of umms cores testifies in favour of the genuine nature of some of these latter. The $H_2$ emission seems to be directly connected with the $mm$ cores in about one third of the investigated fields (namely MMS2-19-22-29, umms16-19 and possibly MMS16), but we are able to identify a possible (NIR) driving source in only two cases (MMS22 and MMS29, see Tab.\,\ref{table:cores-NIR-assoc}, last column). Given the poor sensitivity of MSX and IRAS observations, Class\,0 protostars might already be embedded in the other dust cores, which represent interesting cases to be investigated with SPITZER. All the fields of young clusters (those associated with cores MMS2-4-12-22 and 25) show $H_2$ emission and close to three peaks (MMS2-4-22), all associated with an IRAS source, we have found a $sub-parsec$ scale jet (with extension 0.30, 0.30 and $0.68\,$pc, respectively). Only in one case the jet seems to be emitted by the most luminous object in the field (i.e. the one that contributes the most to the FIR flux in the IRAS bands), in all the other cases the driving source is a less luminous (and less massive) object in the cluster. A similar result was pointed out by Lorenzetti et al. (2002), who, having surveyed a sample of 12 IRAS protostellar candidates in VMR, have detected $H_2$ emission in 5 fields, clearly coming from low-mass objects clustered around the IRAS source, and not from the IRAS source itself. This feature is likely related to a different duration of the jet phase in low and high luminosity (mass) sources: this topic will be discussed in depth in a forthcoming paper (Giannini et al. 2007, in preparation).

\section{Conclusions} \label{par:Conclusions}
A southern sky area of 1$\times$1\,deg$^2$ belonging to the star forming region VMR-D, previously surveyed at $mm$\,wavelengths to identify the dust cores, has been studied by means of both IR catalogues (IRAS/MSX/2MASS) and a set of new dedicated observations, to identify the young protostellar counterparts associated to dust cores. The motivation for the new IR observations is twofold: {\it (i)} to perform a complete survey of all the recognized dust cores in the $H_2$ narrow-band filter (2.12\,$\mu$m) to search evidence of protostellar jets; and {\it (ii)} to obtain a broad band $N$ (10.4\,$\mu$m) survey of those cores associated to embedded clusters, aiming to pick up the source(s) that mainly contribute(s) in the far-IR regime.\\
The main conclusions of this work are summarized here:
\begin{itemize}
\item[-] In the majority of cases, MIR and/or FIR sources associated with dust cores do not coincide with the $mm$ peaks, although they are located in their vicinity. In those cases of close-by cores, often the IR source is located in between them.
\item[-] The resolved $mm$ cores (i.e. those larger than the instrumental beam) are more frequently associated to a NIR or FIR counterpart than the unresolved ones (smaller than the instrumental beam). The existence of signs of star formation activity around these latter in the form of $H_2$ jets, however, attests the genuine nature of lots of them. 
\item[-] The statistics of active vs inactive cores is in good agreement with that found in other star forming regions, but should be critically revised in the light of more sensitive observations that will become available in near future (e.g. SPITZER MIPS and IRAC maps).
\item[-] The SEDs of the associated sources present a slope $\alpha=d\log(\lambda F_{\lambda})/d\log(\lambda)$ between 2 and 10\,$\mu$m whose distribution is strongly peaked at values typical of Class\,I sources ($0<\alpha<3$), in some cases even larger.
\item[-] An attempt has been done to search for Class\,0 sources. Ten IRAS sources (with upper limits at 12\,$\mu$m) do not present any NIR reliable counterpart, but no one of them satisfies the criterion $L_{bol}/L_{1.3mm} \lesssim 2 \times 10^4$ proposed by Andr\'e et al. (1993). However, considered the probable overestimates of the IRAS fluxes (and bolometric luminosities), we indicate the three objects with the lowest $L_{bol}/L_{1.3mm}$ ratios as the youngest IRAS sources of the region.
\item[-] The sources associated to the dust cores, both resolved and unresolved, have all the same FIR colours, typical of a black-body stratification between 50 and 1000\,K, with a stronger contribution of the former component. In other words, sources with FIR colours typical of pre-main sequence T-Tauri and Herbig Ae/Be stars seem to be absent, indicating VMR-D as a young ($\sim\,10^6$\,years) region.
\item[-] Sources associated with unresolved cores are systematically less luminous (average $L_{bol}\,\simeq\,1.5\times 10^2\,L_{\odot}$) than those related to the resolved ones (average $L_{bol}\,\simeq\,1.5 \times 10^3 L_{\odot}$), providing evidence that two modalities of star formation, namely low- and intermediate-mass, are simultaneously present in VMR-D.
\item[-] This occurrence is also confirmed by the existence of a further (low luminosity) population of young objects that have the same colours as the sources associated to the dust cores (resolved or unresolved), but are located, on the contrary, in the diffuse warm dust and along the gas filaments.
\item[-] Observing the dust cores in the $H_2$ 1-0S(1) line evidences a high detection rate (30\,\%) of jets driven by sources located inside the cores. They appear not driven by the most luminous objects in the field, but rather by less luminous (and less massive) objects in surrounding star clusters, testifying the co-existence of both low and intermediate-mass star formation, being the former more likely associated to molecular jet activity. Moreover the jet compactness (at $sub-arcsecond$ scales) is a further indication for the prevalence of  very young objects.
\end{itemize}

%\section{Acknowledgements}

\Online
\clearpage
\appendix \label{Appendix}

\section{Counterparts of the dust cores}
In the following we present a brief discussion of the NIR to FIR associations found for each dust core. Other signs of star formation activity (e.g. $H_2$ knots and jets or young embedded clusters) will be also evidenced. For each core (or group of cores) we will show the SofI $H_2$ (or 2MASS $H$ band) gray scale image supplied with: dust emission contours, position of the most interesting sources, coverage of IRAC2 and Timmi2 fields of view and arrows indicating the occurrence of $H_2$ knots (see Sect.\,\ref{par:MMS1} for more details). Whenever needed for the analysis, the corresponding colour-colour diagrams and SED plots will be shown as well.

\begin{figure*}
   \centering
   \includegraphics[width=10cm]{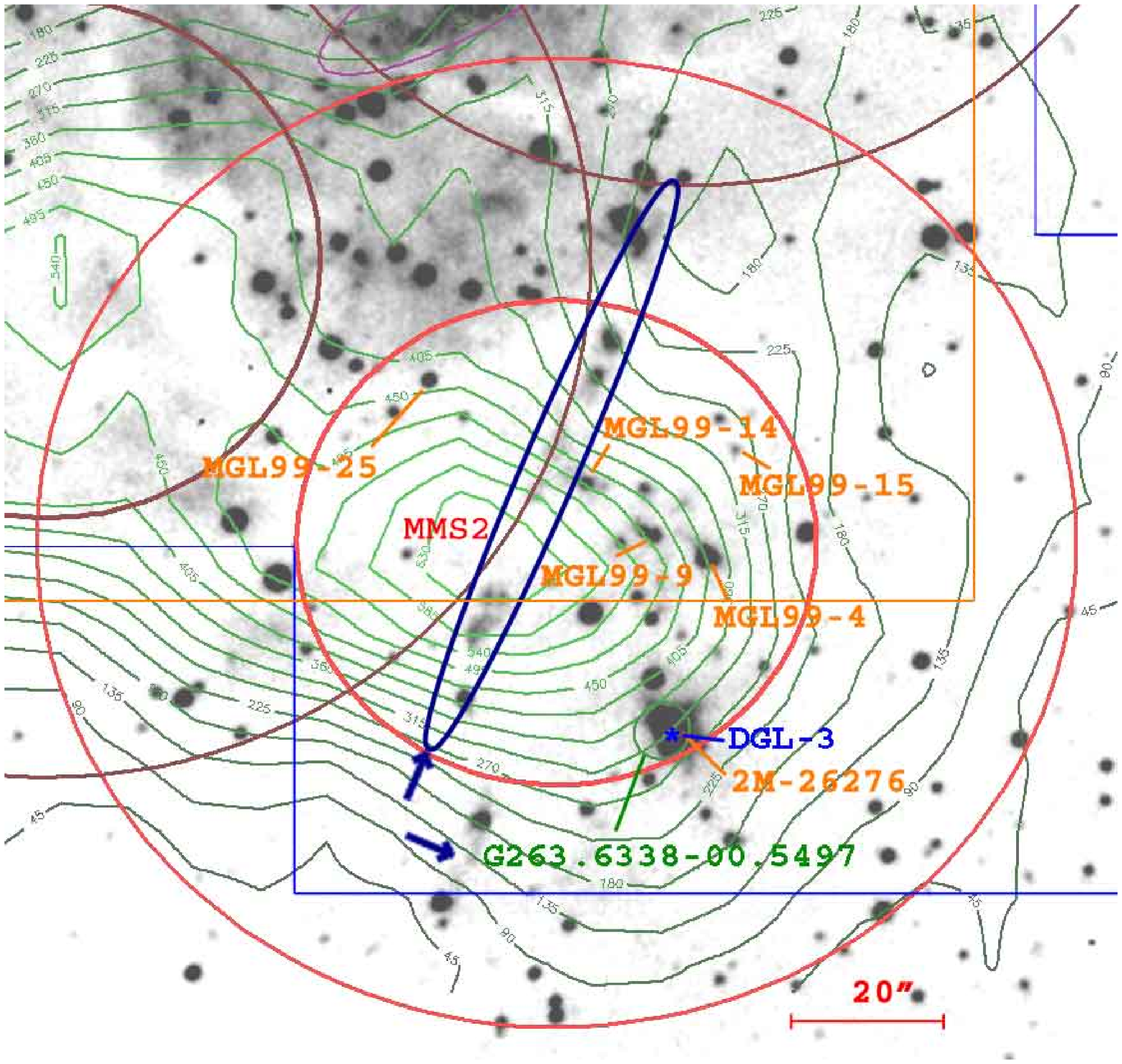}
   \caption{MMS2 field of view (center [J2000]: 08:45:34.200, -43:51:54.40). Grayscale image: $H_2$ emission; green contours: dust continuum (from 3\,$\sigma$, in steps of 3\,$\sigma$); red ellipses: FWHM-ellipse and 2FWHM-ellipse of the core (see Sect.\,\ref{par:FIR-associations} for an explanation); green and magenta ellipses: MSX and IRAS 3\,$\sigma$ positional uncertainty; orange line: IRAC2 field of view.}
              \label{Fig:MMS2}
\end{figure*}

\begin{figure*}
\centering
\includegraphics[width=\textwidth]{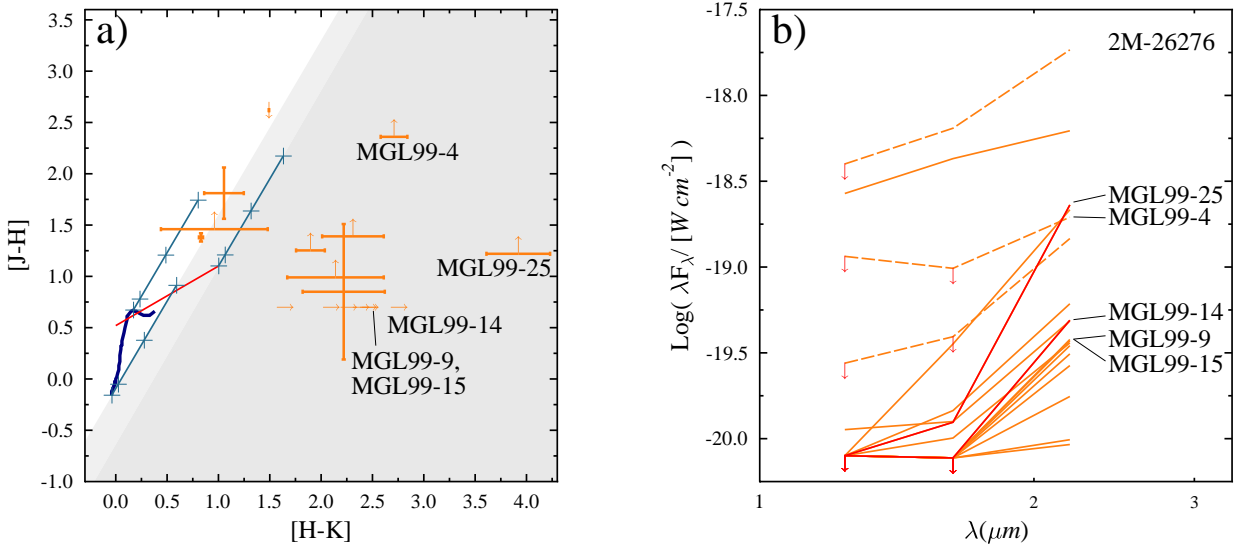}
   \caption{{\it a)} Colour-colour diagram (see text for more details) of the NIR sources within the FWHM-ellipse of MMS2. The Spectral Energy Distributions of the very red ones are shown in panel {\it b)}. To reduce confusion the wavelength range is limited to the $J$, $H$, $K$ bands and the main objects are in red colour. Dashed lines refer to 2MASS sources and arrows denote upper limits.}
              \label{Fig:MMS2-colcol-SED}
\end{figure*}

\begin{figure*}
   \centering
   \includegraphics[width=8cm]{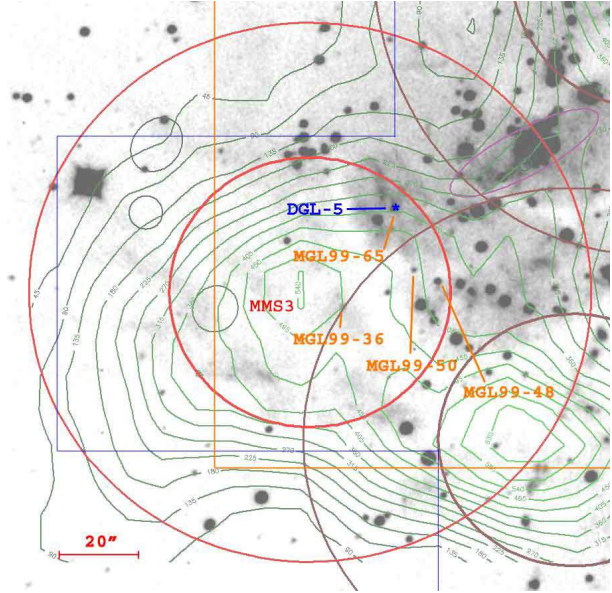}
   \caption{MMS3 field of view (center [J2000]: 08:45:39.5, -43:51:25.0).}
              \label{Fig:MMS3}
\end{figure*}

\begin{figure*}
   \centering
   \includegraphics[width=\textwidth]{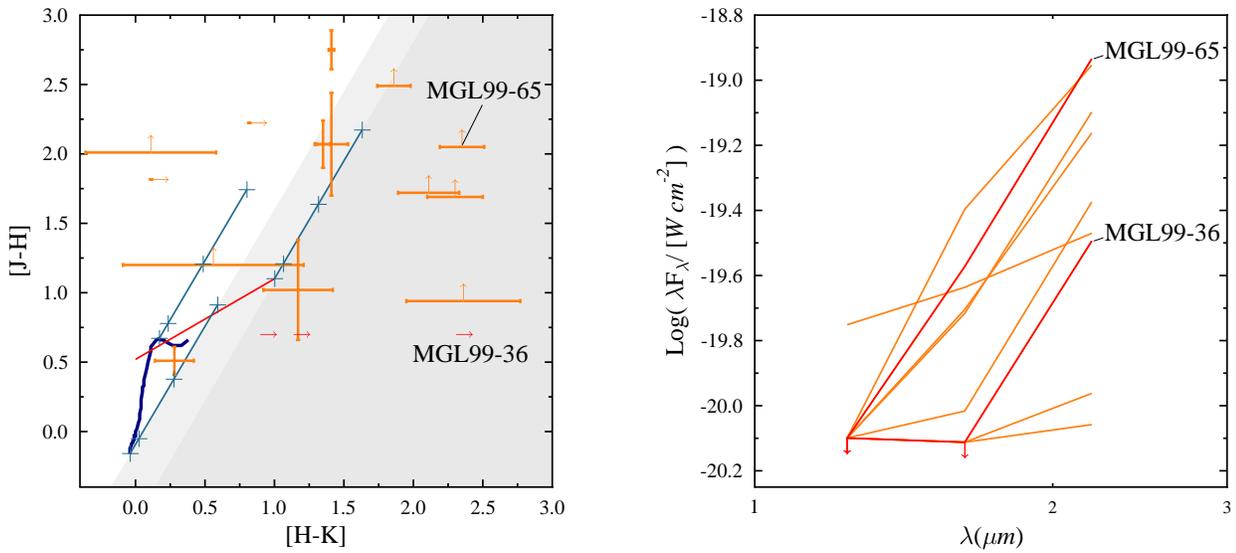}
   \caption{MMS3: colour-colour and SED diagrams.}
              \label{Fig:MMS3-colcol-SED}
\end{figure*}

\begin{figure*}
   \centering
   \includegraphics[width=9cm]{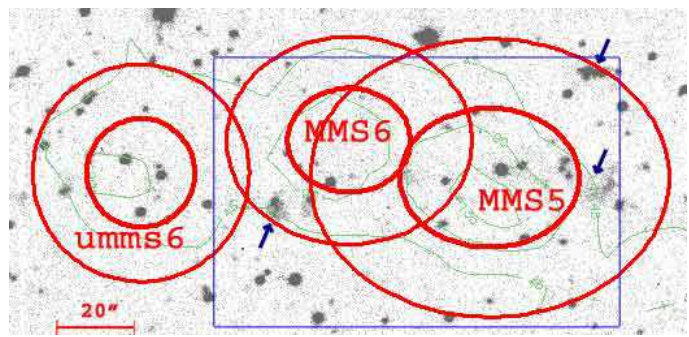}
   \caption{MMS5-6 and umms6 field of view (center [J2000]: 08:46:52.0, -43:53:01.3).}
              \label{Fig:MMS5-6-umms6}
\end{figure*}

\begin{figure*}
   \centering
   \includegraphics[width=8cm]{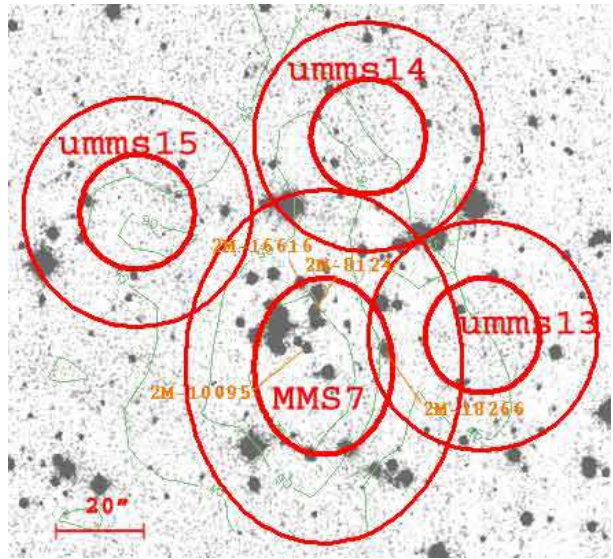}
%             \label{Fig:MMS7-umms13-14-15}
\includegraphics[width=10cm]{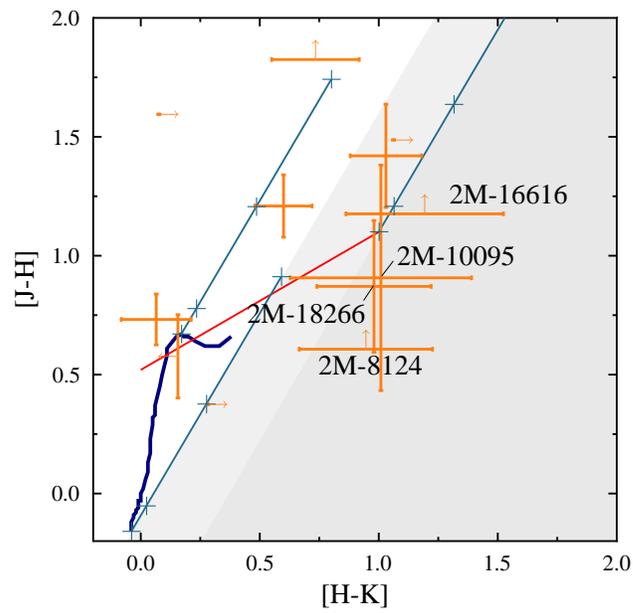}
   \caption{MMS7 and umms13-14-15 field of view (center [J2000]: 08:47:58.9, -43:39:22.9) and colour-colour diagram.}
              \label{Fig:MMS7-umms13-14-15-colcol}
\end{figure*}

\begin{figure*}
   \centering
   \includegraphics[width=10cm]{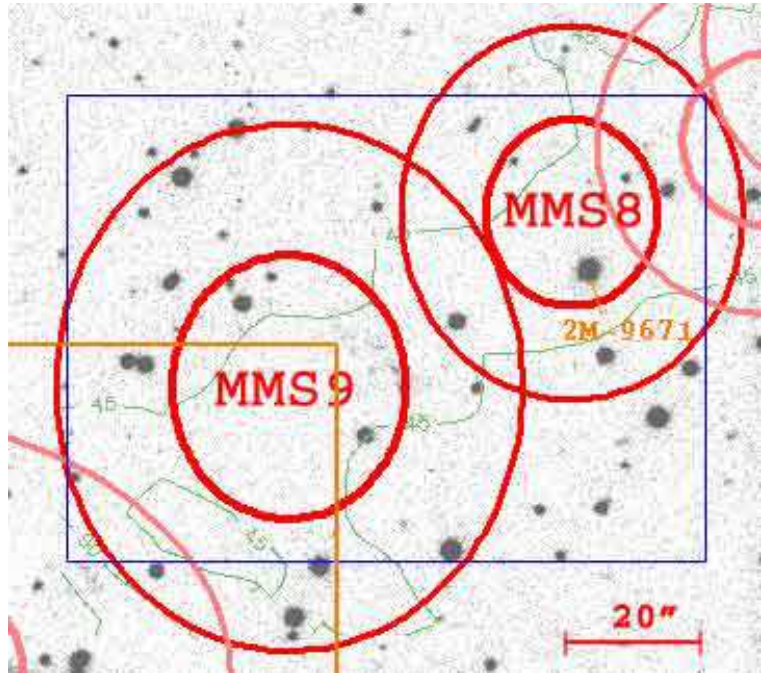}
\includegraphics[width=10cm]{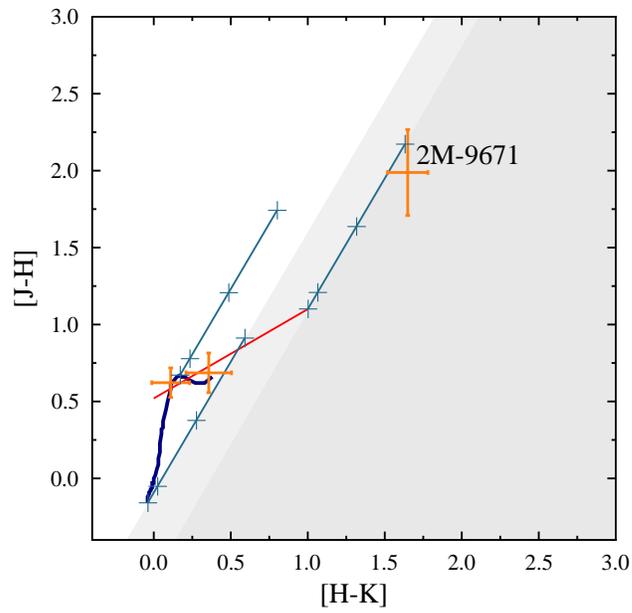}
   \caption{MMS8-9 field of view (center [J2000]: 08:48:41.3, -43:31:36.2) and colour-colour diagram.}
              \label{Fig:MMS8-9}
\end{figure*}

\begin{figure*}
   \centering
   \includegraphics[width=8cm]{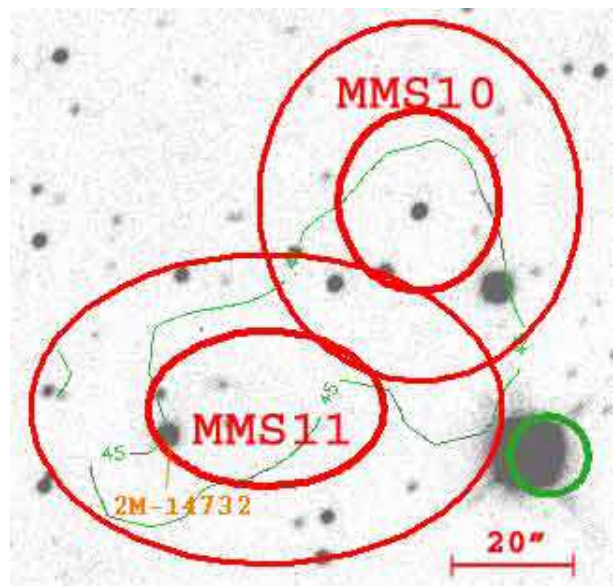}
   \includegraphics[width=10cm]{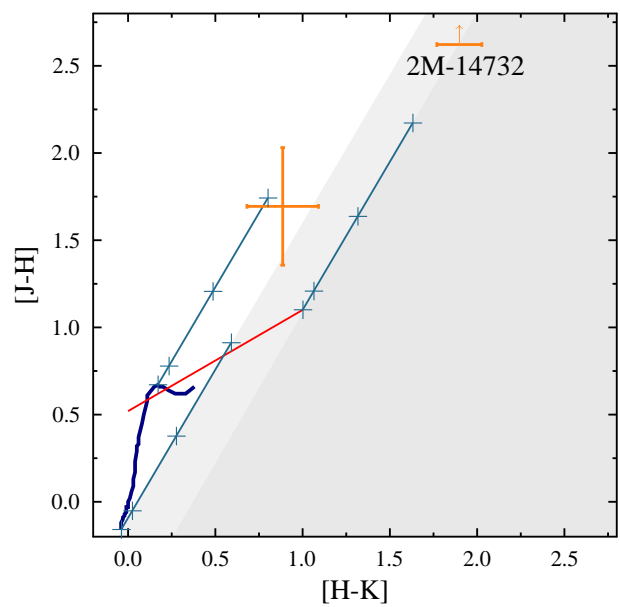}
   \caption{MMS10-11 field of view (center [J2000]: 08:48:44.2, -43:37:19.1) and colour-colour diagram.}
              \label{Fig:MMS10-11}
\end{figure*}

\begin{figure*}
   \centering
   \includegraphics[width=10cm]{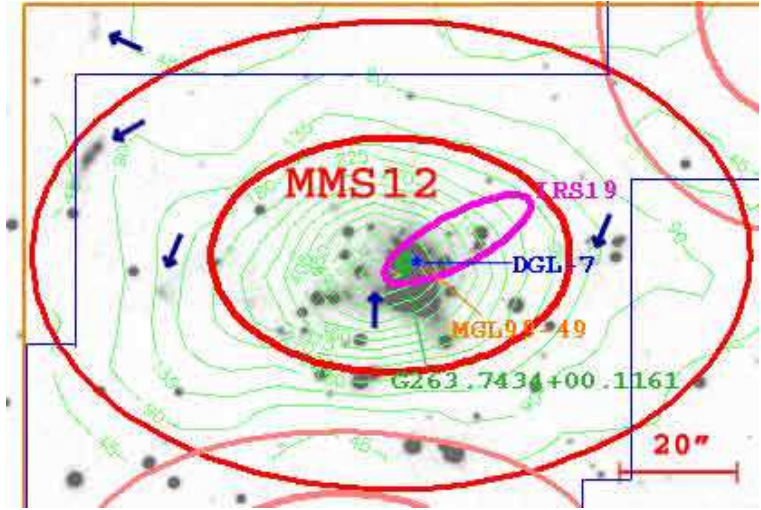}
   \caption{MMS12 field of view (center [J2000]: 08:48:48.5, -43:32:20.8).}
              \label{Fig:MMS12}
\end{figure*}

\begin{figure*}
   \centering
   \includegraphics[width=\textwidth]{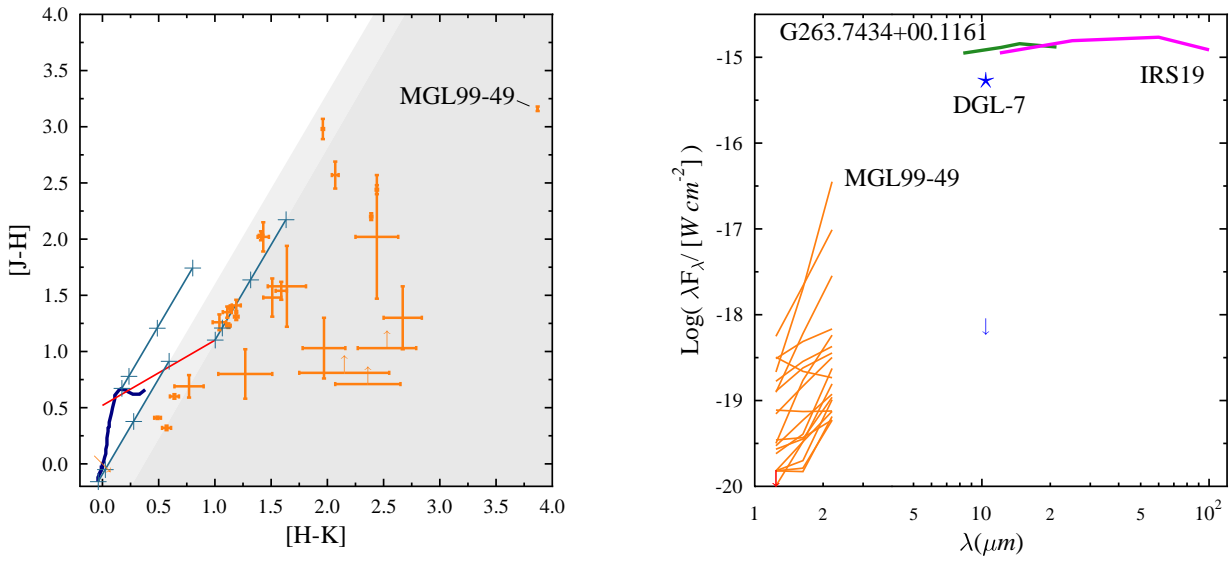}
   \caption{MMS12 colour-colour and SED diagrams.}
              \label{Fig:MMS12-colcol-SED}
\end{figure*}

\begin{figure*}
   \centering
   \includegraphics[width=8cm]{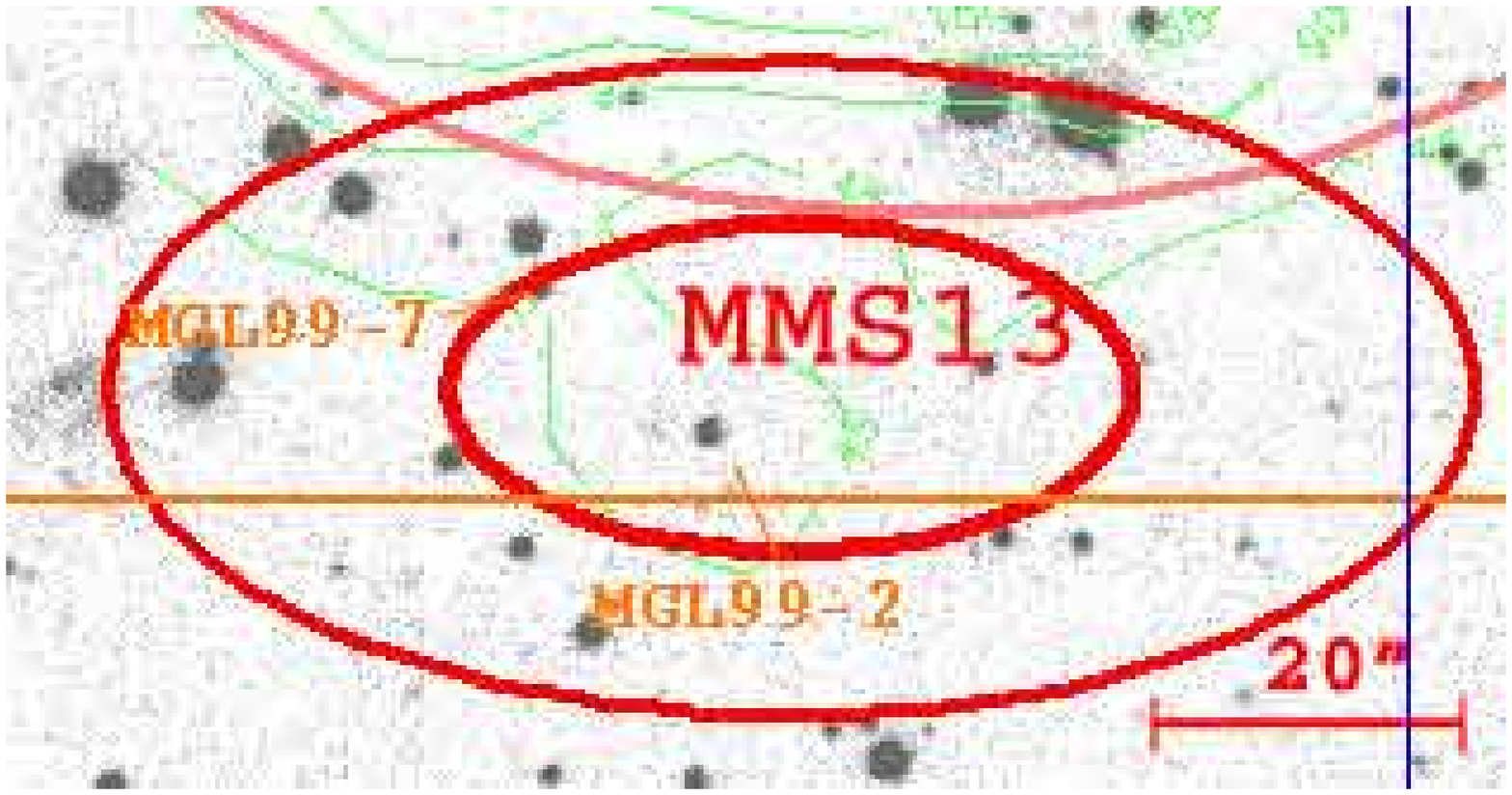}
   \includegraphics[width=8cm]{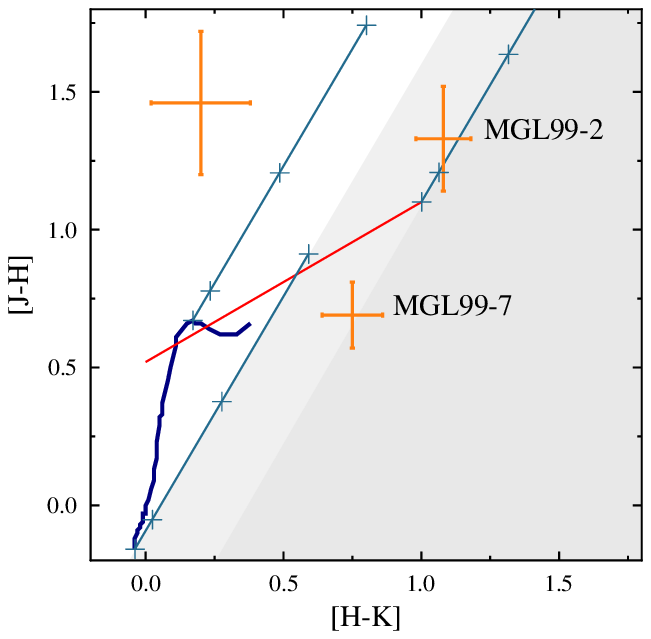}
   \caption{MMS13 field of view (center [J2000]: 08:48:49.4, -43:33:11.2) and colour-colour diagram.}
              \label{Fig:MMS13}
\end{figure*}

\begin{figure*}
   \centering
   \includegraphics[width=8cm]{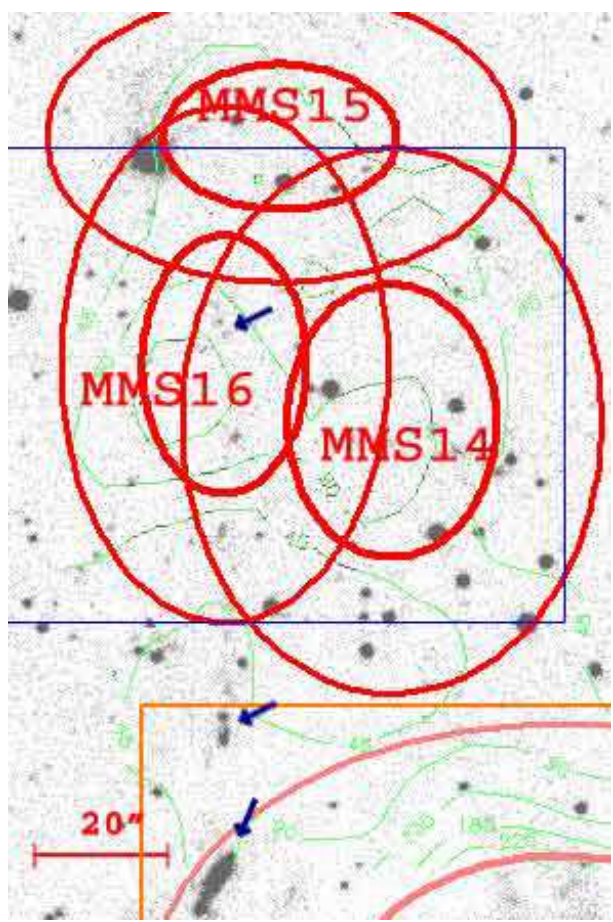}
   \caption{MMS14-15-16 field of view (center [J2000]: 08:48:51.6, -43:31:09.9).}
              \label{Fig:MMS14-15-16}
\end{figure*}

\begin{figure*}
   \centering
   \includegraphics[width=10cm]{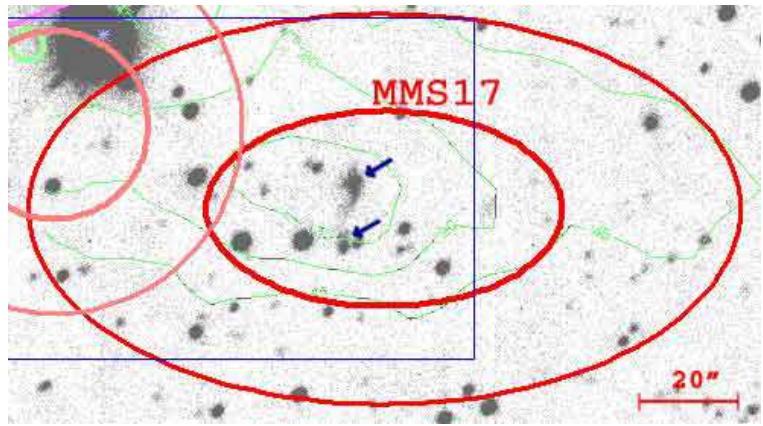}
   \caption{MMS17 field of view (center [J2000]: 08:48:57.2, -43:38:23.1).}
              \label{Fig:MMS17}
\end{figure*}

\begin{figure*}
   \centering
   \includegraphics[width=8cm]{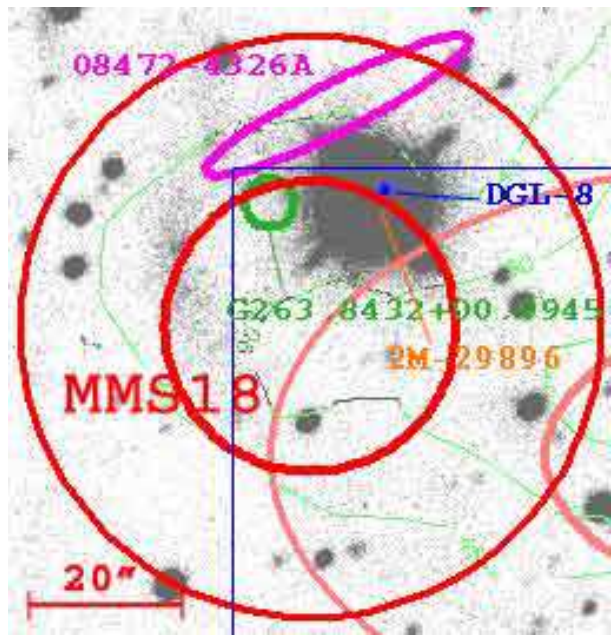}
   \caption{MMS18 field of view (center [J2000]: 08:49:03.2, -43:38:05.2).}
              \label{Fig:MMS18}
\end{figure*}

\begin{figure*}
   \centering
   \includegraphics[width=\textwidth]{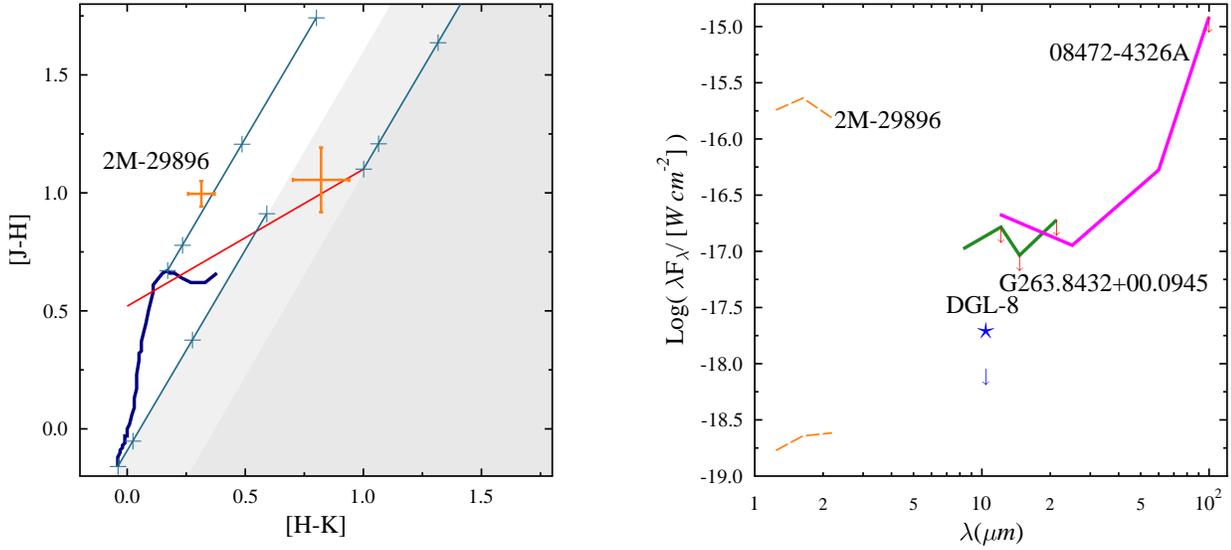}
   \caption{MMS18 colour-colour and SED diagrams.}
              \label{Fig:MMS18-colcol-SED}
\end{figure*}

\begin{itemize}
\item{\textbf{MMS2} (Figs.\,\ref{Fig:MMS2}, \ref{Fig:MMS2-colcol-SED}):} one MSX associated object inside the FWHM-ellipse, G263.6338-00.5497, with a $N$ band counterpart, DGL 3, having a flux ($F_{10.4\mu m}=0.25\,$Jy) smaller than that measured by MSX ($F_{8.3\mu m}=0.4\,$Jy).

Inside the FWHM-ellipse, many NIR stars (recognizable in the SofI $H_2$ image) have not been detected by 2MASS and are only upper limits for IRAC2 in both $J$ and $H$ bands. The object showing the highest colour excess (MGL99 25) and steepest spectral index (Fig.\,\ref{Fig:MMS2-colcol-SED}) is not visible in the $N$ band (at 30\,mJy sensitivity level). We have tentatively indicated that one in Tab.\,\ref{table:cores-NIR-assoc} as candidate NIR counterpart. However, an intense $H_2$ line emission is present all over the core, both diffuse and in knots and, remarkably, a well collimated $H_2$ jet crosses the very center of the core. No reliable exciting source has been detected along the jet. Likely, it is heavily embedded near the peak position and contributes significantly (or mainly) to the observed dust emission.\\

\item{\textbf{MMS3} (Figs.\,\ref{Fig:MMS3}, \ref{Fig:MMS3-colcol-SED}):} an IRAS source (IRS16) is marginally associated to the peak (see Sect.\,\ref{par:MMS1}). Three MSX detections (one of them inside the FWHM-ellipse) fall in a region of enhanced and diffuse 8\,$\mu$m emission and don't seem to be point-like (as suggested by the MSX image and by the failed detection in $N$). A Timmi2 source (DGL 5, identified with MGL99 65) is instead observed in the direction of the IRAS source and NIR cluster.

A steep decrease in the number of NIR detections towards the peak suggests a high extinction level. Moreover, the majority of the IRAC2 sources within the FWHM-ellipse presents very red colours. MGL99 36, the one closest to the peak position, could be the main source associated to the millimeter core.\\

\item{\textbf{MMS4}:} a detailed analysis of this core has been already presented in a dedicated paper \cite{Giannini2005}, which the reader is referred to.\\

\item{\textbf{MMS5-6, umms6} (Fig.\,\ref{Fig:MMS5-6-umms6}):} a dust elongated structure of connected cores pointing towards south-west in the direction of MMS4, the brightest core of the whole dust map; a dust filament which is likely undergoing fragmentation. Neither MIR-FIR associations, nor interesting NIR sources are present (although no IRAC2 data are available); signs of star formation activity are quite hidden and can be only recognized as a faint $H_2$ emission around MMS5 and MMS6.\\

\item{\textbf{MMS7, umms13-14-15} (Fig.\,\ref{Fig:MMS7-umms13-14-15-colcol}):} resolved core (MMS7) surrounded by three unresolved cores (umms13-14-15). The $J$, $H$ and $K_s$ photometry of the cluster of about 20 members around the central core will be analyzed in a forthcoming paper. Neither FIR emission nor $H_2$ features are present (no 10\,$\mu$m image has been collected). We just remark here the colour excess of 4 very red stars within the FWHM-ellipse of MMS7.\\

\item{\textbf{MMS8-9} (Fig.\,\ref{Fig:MMS8-9}):} these cores are part of a long tail of connected cores (extending for about 1\,pc), going from the bright core MMS12 to umms19, whose dynamical behaviour is not clear \cite{Massi2007}. No FIR point sources or clues of $H_2$ emission have been observed. Timmi2 observations, although not covering the whole dust emission, gave negative results as well. Only one very red NIR object, 2M 9671 (no IRAC2 data available), lies within the FWHM-ellipse of MMS8.\\

\item{\textbf{MMS10-11} (Fig.\,\ref{Fig:MMS10-11}):} without any MIR-FIR counterpart, these are the only two resolved cores included in the $^{13}$CO map that lack of an associated CO clump \cite{Elia2007}. The colour-colour diagram of the 2MASS sources (no IRAC2 data available) within the FWHM-ellipse gives only one reddened candidate (2M-14732) for MMS11.\\

\item{\textbf{MMS12} (Figs.\,\ref{Fig:MMS12}, \ref{Fig:MMS12-colcol-SED}):} one of the brightest cores, MMS12 ($\sim 18 M_{\odot}$) coincides with a young embedded cluster having in its center the IRAS source 08470-4321 (IRS19), the MSX G263.7434+00.1161 and the Timmi2 DGL 7 object. The new observed flux at 10\,$\mu$m ($F_{10.4\mu m}=18.3\,$Jy) is less than a half of the IRAS/MSX measurements at 8-12\,$\mu$m. We ascribe this discrepancy to the presence of a strong diffuse contribution to the flux at these wavelengths.\\
The correspondence of these sources with the NIR object MGL99 49 has been already discussed in Massi et al. (1999) and is confirmed here.\\
Complex and intense $H_2$ emission is also present, especially at the peak position, and at the eastern part of the core; it cannot be clearly distinguished as a single jet-like structure.\\

\item{\textbf{MMS13} (Fig.\,\ref{Fig:MMS13}):} no MIR-FIR association and no significant $H_2$ emission for this core at the south of MMS12. The two possible NIR counterparts have red colours (MGL99 2, MGL99 7).\\

\item{\textbf{MMS14-15-16} (Fig.\,\ref{Fig:MMS14-15-16}):} neither MIR-FIR associations nor NIR red sources detected by 2MASS (no IRAC2 data). We note a faint $H_2$ emission aligned with both the peak MMS16 and the $H_2$ knot visible at the east of MMS12 (see also Fig.\,\ref{Fig:MMS12}). We cannot exclude the existence of an embedded exciting source near the MMS16 peak position.\\

\item{\textbf{MMS17} (Fig.\,\ref{Fig:MMS17}):} a double $H_2$ knot (probably jet-like) in proximity of the peak suggests the existence of star forming activity, but no MIR-FIR objects have been detected and the 2MASS data do not point out any interesting source within the FWHM-ellipse.\\

\item{\textbf{MMS18} (Figs.\,\ref{Fig:MMS18}, \ref{Fig:MMS18-colcol-SED}):} one IRAS source (08472-4326A) is associated within 2FWHM-ellipse, while one MSX source (G263.8432+00.0945) and one Timmi2 object (DGL 8) are inside the FWHM-ellipse. The positional uncertainties of these three objects seem to exclude their coincidence, although the IRAS and MSX fluxes (due to their beam sizes) are surely contaminated by the Timmi2 source (whose counterpart, 2M 29896, peaks in the $H$ band) and probably by diffuse emission, visible in the $H_2$ filter as well.\\

\item{\textbf{MMS19} (Fig.\,\ref{Fig:MMS19}):} core connected to MMS20 and MMS21. Neither MIR-FIR sources associated nor $H_2$ emission detected. We signal two very red 2MASS objects (2M-36076, 2M 36192) within the FWHM-ellipse.\\

\item{\textbf{MMS20} (Fig.\,\ref{Fig:MMS20}):} one IRAS (08474-4323, corresponding to MSX G263.8221+00.1494) source turns out to be very marginally associated (positional uncertainty tangent to the 2FWHM-ellipse), but the dust emission around its position seems to be negligible.
One Timmi2 source (DGL 9) has been detected within 2FWHM-ellipse, in the middle between this core and MMS21 not signalled by IRAS/MSX (SofI images points out at least three objects, partially resolved by 2MASS in two very red sources, 2M 27831 and 2M 37241).\\

\item{\textbf{MMS21} (Figs.\,\ref{Fig:MMS21}, \ref{Fig:MMS21-colcol-SED}):} one IRAS source, 08474-4325, whose fluxes indicate this one as one of the youngest objects of VMR-D, coincides with the peak position. The Timmi2 observation has given no results (for the source DGL 9 see MMS20 description) and no MSX sources are reported in the catalogs.
Three 2MASS sources fall close to the IRAS uncertainty ellipse center, two of which (2M-29953, 2M 37193) have very red colours. The SofI $H_2$ image, however, points out the presence of a complex morphology, which cannot be resolved in individual sources with the SofI spatial resolution.\\

\item{\textbf{MMS22} (Figs.\,\ref{Fig:MMS22}, \ref{Fig:MMS22-colcol-SED}):} characterized by a powerful bipolar jet (0.7 pc long, discussed in a forthcoming paper) arising from the IRAS source 08476-4306 (IRS20), located 10$^{\prime \prime}$ at the west of the peak. In correspondence with the IRAS source there are a young cluster, a MSX point source (G263.6177+00.3652) and two 10\,$\mu$m objects: DGL 11 (main counterpart of the IRAS/MSX object) and DGL 10 (2\,$\sigma$ detection, outside the FWHM-ellipse). The DGL 11 flux, $F_{10.4\mu m}=2.01\,$Jy, is significantly smaller than those measured by MSX and IRAS: $F_{8.3\mu m}=3.9\,$Jy, $F_{12\mu m}=5.7\,$Jy, $F_{12.1\mu m}=6.0\,$Jy, but the Timmi2 observation points out a diffuse emission which can have contributed to the MIR fluxes measured by IRAS and MSX. The presence of nebular NIR and $H_2$ emission completes the picture of this crowded region. The NIR cluster characteristics have been discussed in detail by Massi et al. (1999) and the conclusions reported in that paper about the candidate NIR counterpart (MGL99 98) of IRS20 are confirmed by our Timmi2 observation. MGL99 98 is also the best candidate as exciting source of the jet, although the complexity of the jet morphology and the difficulty to discriminate between nebular and point-like unresolved emission makes this identification questionable. We also remark one dark strip clearly visible in the $H_2$ image, likely due to an obscuring dust lane crossing the cluster center.\\

\item{\textbf{MMS23} (Fig.\,\ref{Fig:MMS23})}: isolated, small core with no IR detected sources near the peak (not observed at 10\,$\mu$m). $H_2$ multiple knots are visible to the north of the peak together with a faint emission near the center. We cannot exclude the presence of an embedded, young, low-mass stellar object producing that emission, but more sensitive observations are required to confirm this possibility.\\

\item{\textbf{MMS24} (Fig.\,\ref{Fig:MMS24})}:
this core is connected with the brighter MMS22 and is characterized by diffuse $K$ emission close to a MSX source (within 2FWHM-ellipse) not detected by Timmi2. The NIR source MGL99 90, at the border of the FWHM-ellipse, shows the highest colour excess.\\

\item{\textbf{MMS25-26} (Figs.\,\ref{Fig:MMS25-26}, \ref{Fig:MMS25-26-colcol-SED})}: this complex region is constituted by two not well resolved cores (also linked to the brighter MMS27) and, between them, in correspondence with a young NIR cluster \cite{Massi2006}, there are the IRAS source 08477-3459 (IRS21), the MSX source G264.3225-00.1857 and the Timmi2 object DGL 12 (corresponding to the NIR MGL99 27). Moreover, the 10\,$\mu$m emission observed by Timmi2 shows a diffuse emission (in the surroundings of the source MGL99 35) and, probably, another point source corresponding to MGL99 32, although an artifact in the Timmi2 image prevented us to give a reliable estimate of its flux. Remarkable is also a shell-like $K$ emission (clearly visible also in the $H_2$ image) approximately centered near MGL99 63.\\
The colour-colour diagram (Fig.\,\ref{Fig:MMS25-26-colcol-SED}) points out the existence of many red and very red sources within the FWHM-ellipse of both the cores, lots of which having upper limits in the $J$ and $H$ bands. The SEDs of the MIR-FIR objects and of the NIR stars with larger colour excess show significant discrepancies among the fluxes of the MIR-FIR detections. No clear evidence has been found for NIR stars (if any) that can be most likely associated to the $mm$ core\footnote{The lack of any Timmi2 detectable flux in correspondence of MGL99 50, the previously hypothesized NIR counterpart of the IRAS source \cite{Massi1999}, makes that association questionable. Observing its SED, indeed, it is quite unlikely that it could be missed, if point-like, at 10\,$\mu$m.}.\\

\item{\textbf{MMS27} (Fig.\,\ref{Fig:MMS27})}: although this is one of the brightest cores of the whole dust map, no IRAS-MSX point sources have been detected. Only one faint object (DGL 13), counterpart of the NIR very red 2M 53071 (we lack of IRAC2 data for this source), can be seen at 10 $\mu$m, inside the FWHM-ellipse, together with three more very red sources.

Interesting is the case of 2M 47136, the closest one to the peak: no point-like source can be extracted from the $K$ image which shows instead a very diffuse emission well observable also in the $H_2$ image of Fig.\,\ref{Fig:MMS27}. It is crossed by a dark horizontal strip (quite likely due to obscuring dust). A knot of $H_2$ emission is also visible above this strip and one more on the other side of the core, in the south-west direction, making this core of a peculiar interest for future investigations.\\

\item{\textbf{MMS28-29} (Fig.\,\ref{Fig:MMS28-29}):} inside the FWHM of MMS28 there are the IRAS 08483-4305 and MSX G263.6909+00.4713 sources and, remarkably, $H_2$ aligned knots are visible between the two peaks. We lack of both Timmi2 and IRAC2 observations, but the 2MASS data reveal one very red source, 2M 36339, incompatible with the position of the FIR sources, but aligned with the knots, which could be their exciting source.\\

\end{itemize}

In the following we present the unresolved cores not previously discussed.

\begin{figure*}
   \centering
   \includegraphics[width=8cm]{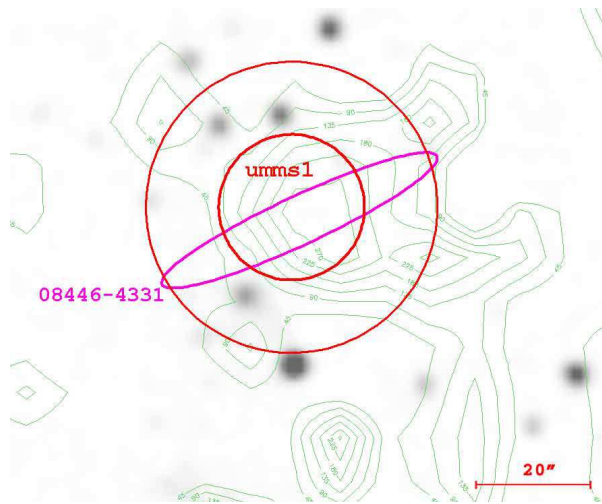}
   \caption{umms1 field of view (center [J2000]: 08:46:25.7, -43:42:28.3) (2MASS $H$ band image).}
              \label{Fig:umms1}
\end{figure*}

\begin{figure*}
   \centering
   \includegraphics[width=8cm]{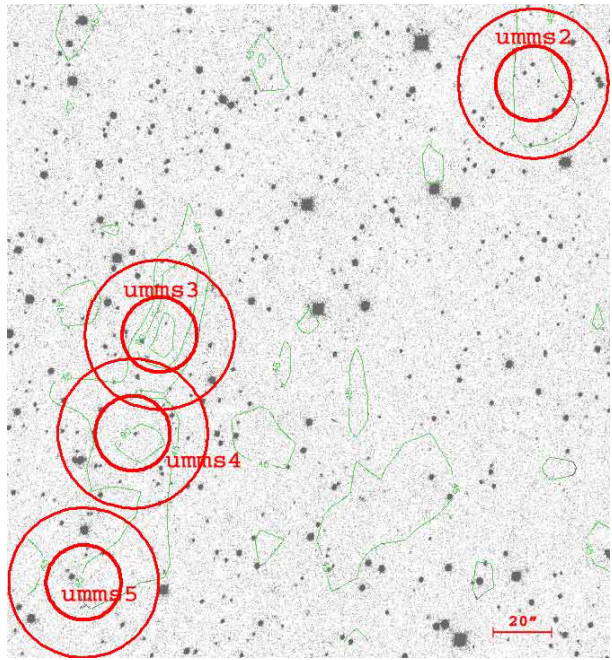}
   \caption{umms2-3-4-5 field of view (center [J2000]: 08:46:44.0, -43:19:48.2).}
              \label{Fig:umms2-3-4-5}
\end{figure*}

\begin{figure*}
   \centering
   \includegraphics[width=8cm]{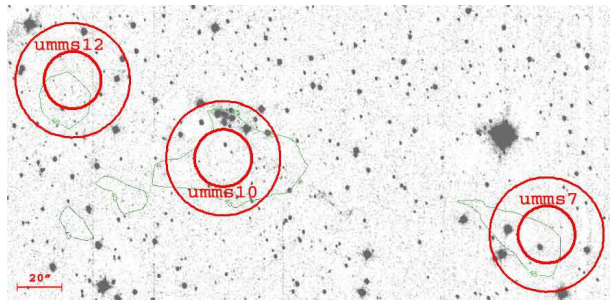}
   \caption{umms7-10-12 field of view (center [J2000]: 08:47:37.4, -43:26:21.7).}
              \label{Fig:umms7-10-12}
\end{figure*}

\begin{figure*}
   \centering
   \includegraphics[width=8cm]{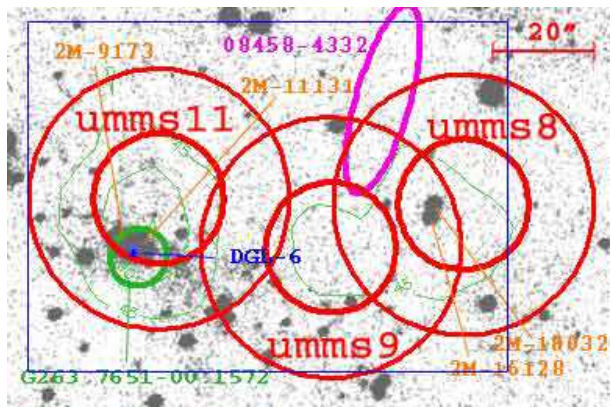}
   \caption{umms8-9-11 field of view (center [J2000]: 08:47:39.6, -43:43:36.1).}
              \label{Fig:umms8-9-11}
\end{figure*}

\begin{figure*}
   \centering
   \includegraphics[width=\textwidth]{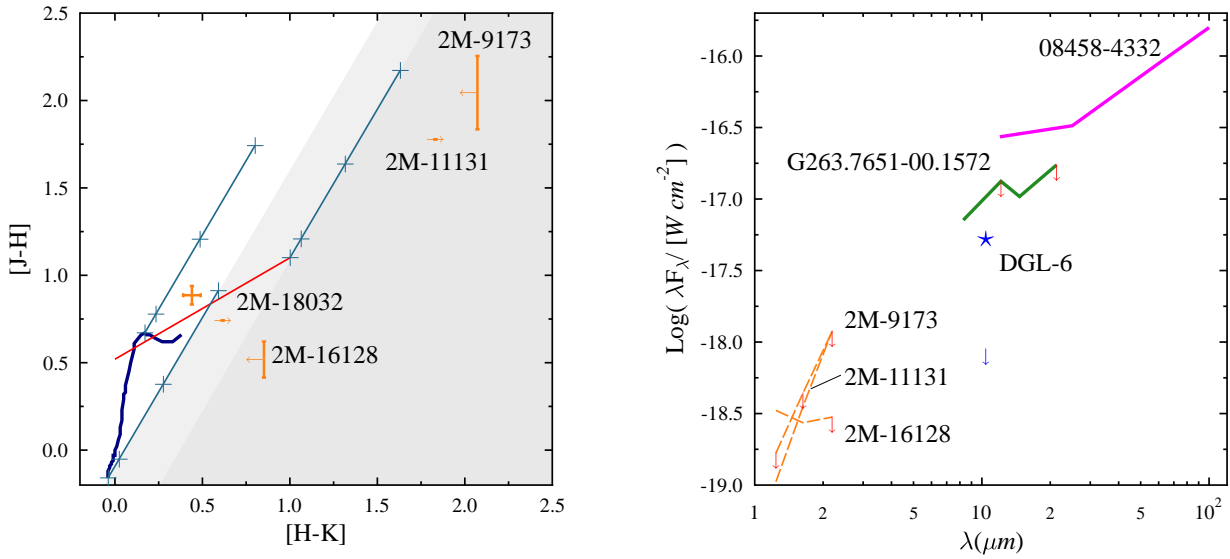}
   \caption{umms8-9-11 colour-colour and SED diagrams.}
              \label{Fig:umms8-9-11-colcol-SED}
\end{figure*}

\begin{figure*}
   \centering
   \includegraphics[width=6cm]{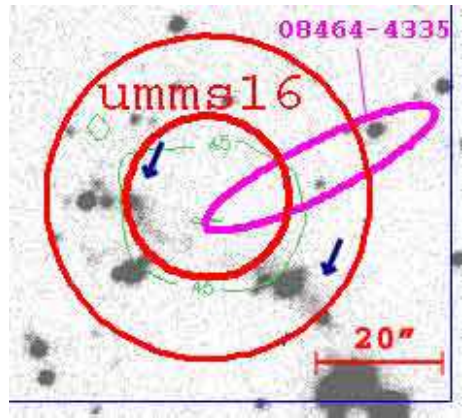}
   \caption{umms16 field of view (center [J2000]: 08:48:15.3, -43:47:06.5).}
              \label{Fig:umms16}
\end{figure*}

\begin{figure*}
   \centering
   \includegraphics[width=8cm]{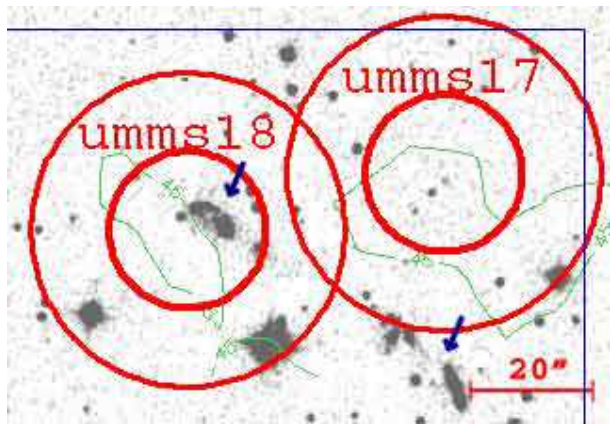}
   \caption{umms17-18 field of view (center [J2000]: 08:48:24.6, -43:31:36.7).}
              \label{Fig:umms17-18}
\end{figure*}

\begin{figure*}
   \centering
   \includegraphics[width=6cm]{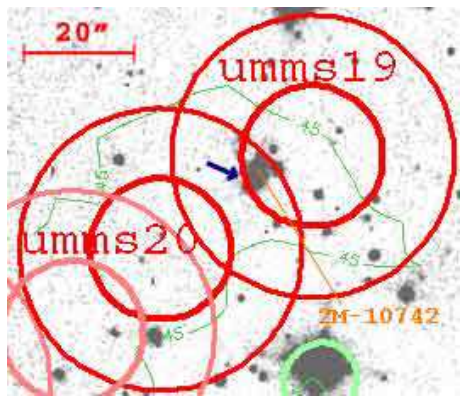}
   \caption{umms19-20 field of view (center [J2000]: 08:48:33.9, -43:30:46.0).}
              \label{Fig:umms19-20}
\end{figure*}

\begin{figure*}
   \centering
   \includegraphics[width=6cm]{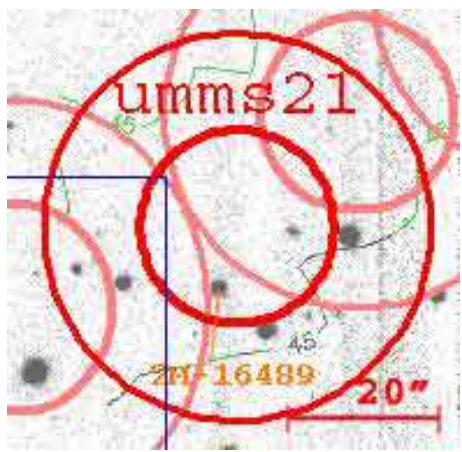}
   \caption{umms21 field of view (center [J2000]: 08:48:36.4, -43:31:11.5).}
              \label{Fig:umms21}
\end{figure*}

\begin{figure*}
   \centering
   \includegraphics[width=6cm]{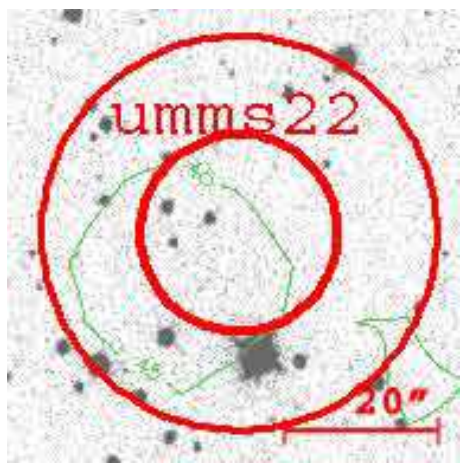}
   \caption{umms22 field of view (center [J2000]: 08:48:36.3, -43:16:45.9).}
              \label{Fig:umms22}
\end{figure*}

\begin{figure*}
   \centering
   \includegraphics[width=7cm]{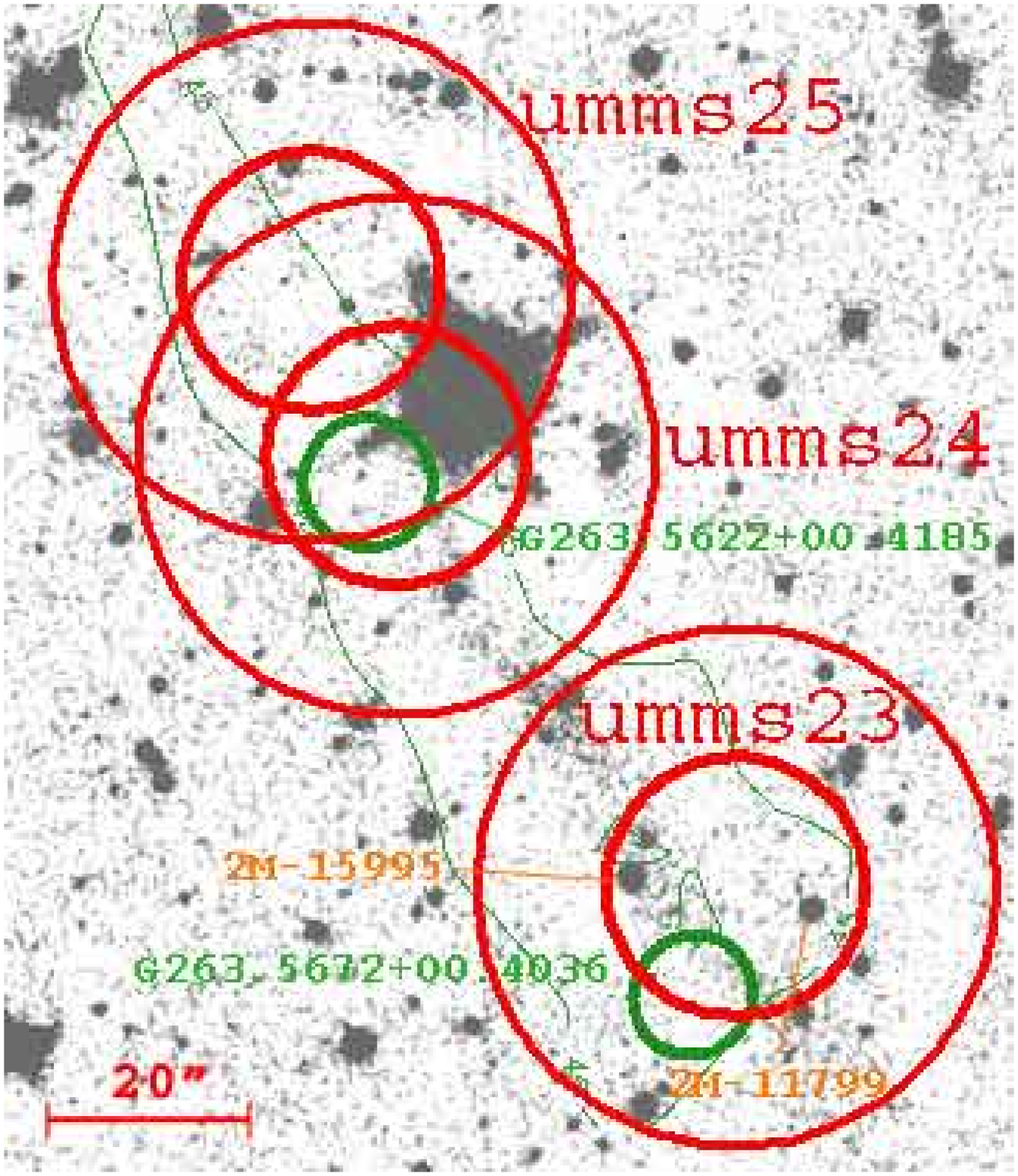}
\includegraphics[width=8cm]{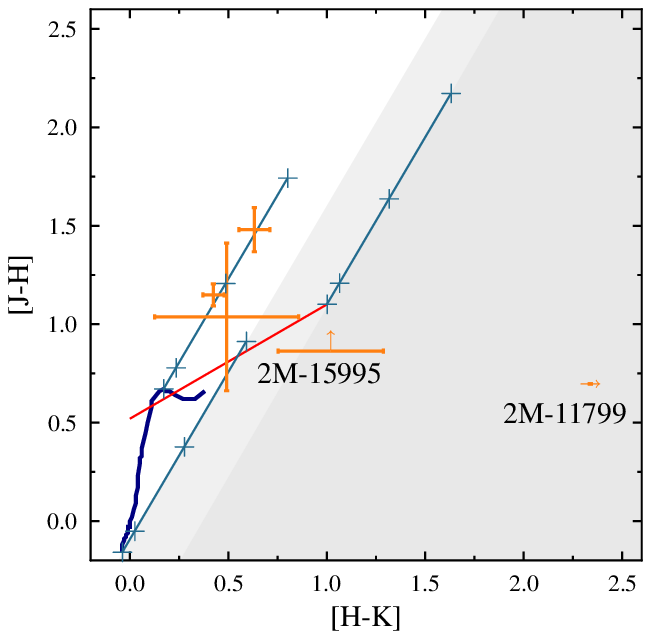}
   \caption{umms23-24-25 field of view (center [J2000]: 08:49:25.9, -43:12:39.2) and colour-colour diagram.}
              \label{Fig:umms23-24-25}
\end{figure*}

\begin{figure*}
   \centering
   \includegraphics[width=6cm]{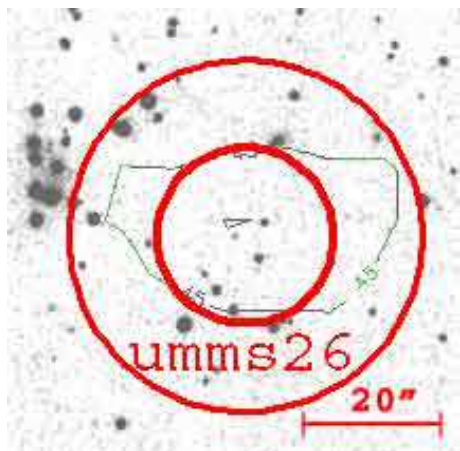}
   \caption{umms26 field of view (center [J2000]: 08:49:58.9, -43:22:55.1).}
              \label{Fig:umms26}
\end{figure*}

\begin{itemize}

\item {\textbf{umms1} (Fig.\,\ref{Fig:umms1}):} despite the high noise level of the dust map in this position, this unresolved core is one of the most interesting cases: {\it (i)} it is by far the most intense core among the unresolved ones, {\it (ii)} its coordinates coincide with an IRAS source showing fluxes increasing with wavelength and {\it (iii)} the $^{12}$CO integrated emission map presents an increase towards this peak, altough it falls immediately outside that map (see note\,\ref{note:umms1-CO}). Unfortunately we lack of both IRAC2 and SofI images at this position and only one 2MASS (red) object falls inside the FWHM-ellipse.\\

\item {\textbf{umms2-3-4-5} (Fig.\,\ref{Fig:umms2-3-4-5}):} region of high noise level of the dust map. These cores are probably artifacts of the reconstruction algorithm and do not present any feature suggesting star formation activity.\\

\item {\textbf{umms7-10-12} (Fig.\,\ref{Fig:umms7-10-12}):} a small cluster within 2FWHM-ellipse of umms10 (forthcoming dedicated paper) is the only noticeably feature of this field.\\

\item {\textbf{umms8-9-11} (Figs.\,\ref{Fig:umms8-9-11}, \ref{Fig:umms8-9-11-colcol-SED}):} dust emission associated with an IRAS source (08458-4332, having flux increasing with $\lambda$, associated to umms8-9) and a MSX-Timmi2 object (G263.7651-00.1572-DGL 6, associated with umms11 and the 2MASS objects 2M 9173 and 2M 11131). One very red and one red NIR object (2M-116128 and 2M 18032, respectively) within the FWHM-ellipse of umms8.\\

\item {\textbf{umms16} (Fig.\,\ref{Fig:umms16}):} core remarkably crossed by $H_2$ jet-like emission and with an IRAS source (08464-4335), although the Timmi2 observation gave no results and no red or very red NIR stars have been observed by 2MASS as possible jet exciting source.\\

\item {\textbf{umms17-18} (Fig.\,\ref{Fig:umms17-18}):} also this couple of cores presents an intense $H_2$ jet-like emission, but without any NIR-MIR interesting source.\\

\item{\textbf{umms19-20} (Fig.\,\ref{Fig:umms19-20}):} these cores are part of a chain of connected cores (see description of MMS8-9). No FIR-MIR %[controllare se oss. da Timmi2] 
point sources have been observed and only a knot of $H_2$ emission is visible in the FWHM-ellipse of umms19, close to a red 2MASS source (2M-10742).\\

\item{\textbf{umms21} (Figs.\,\ref{Fig:umms21}):} connected to the previous cores. No FIR-MIR 
%[controllare se oss. da Timmi2] 
point sources or clues of $H_2$ emission have been observed. One very red NIR object (2M-16489) within the FWHM-ellipse.\\

\item{\textbf{umms22} (Figs.\,\ref{Fig:umms22}):} No FIR point source or clues of $H_2$ emission have been observed (no Timmi2 data available). Absence also of NIR red objects within the FWHM-ellipse.\\

\item{\textbf{umms23-24-25} (Figs.\,\ref{Fig:umms23-24-25}):} thin, elongated dust structure with three cores coinciding with a similar shaped $K$ emission. Two MSX sources, G263.5672+00.4036 and G263.5622+00.4185, are tightly associated to umms23 and umms24, respectively. Two very red sources fall within the FWHM of umms23, but they seem to be incompatible with the position of the MSX sources (for which we suspect a high contamination from diffuse emission).\\

\item{\textbf{umms26} (Figs.\,\ref{Fig:umms26}):} a small cluster outside 2FWHM-ellipse (forthcoming paper) is the only noticeably feature of this field.\\

\end{itemize}

%\clearpage



\begin{figure*}
   \centering
   \includegraphics[width=6cm]{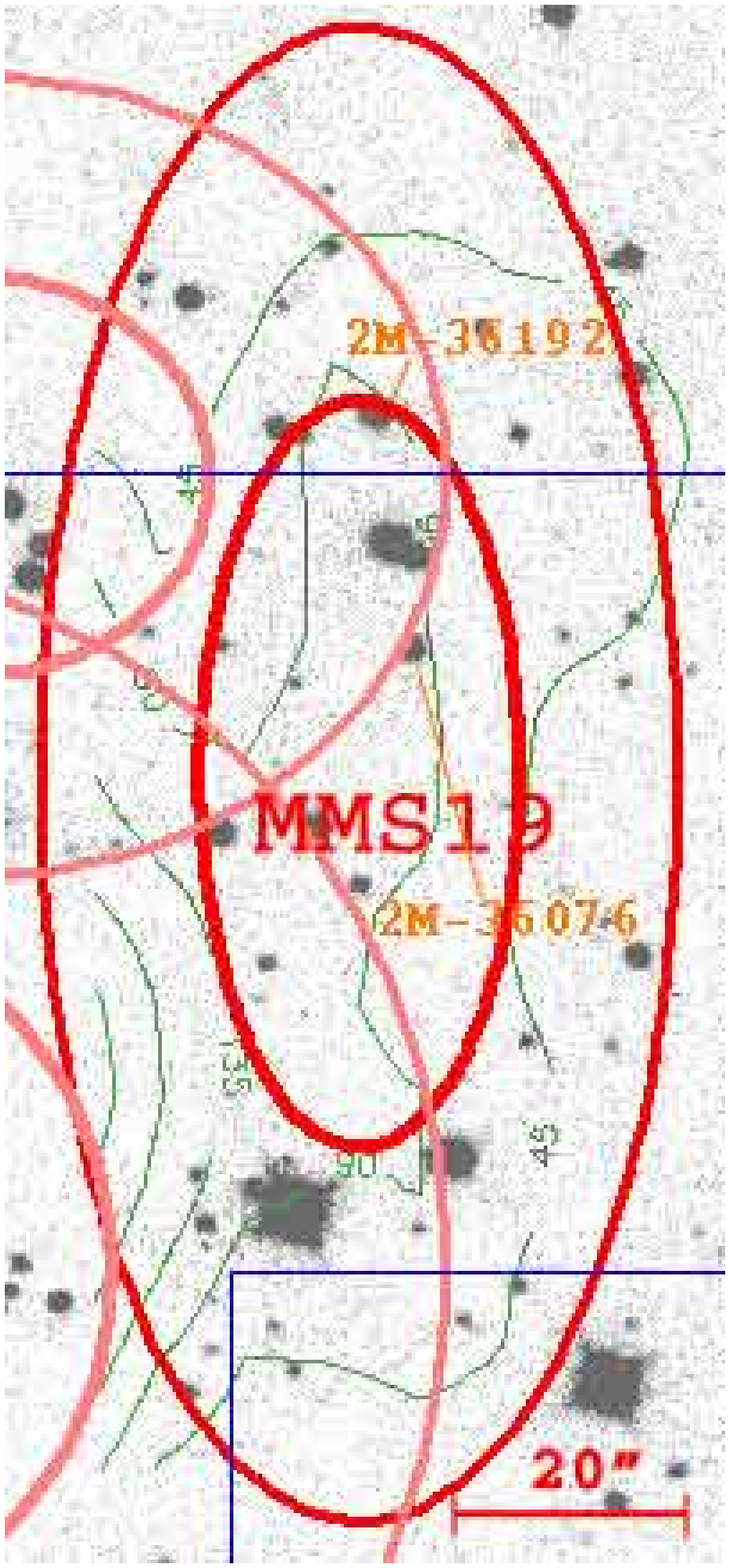}
   \includegraphics[width=8cm]{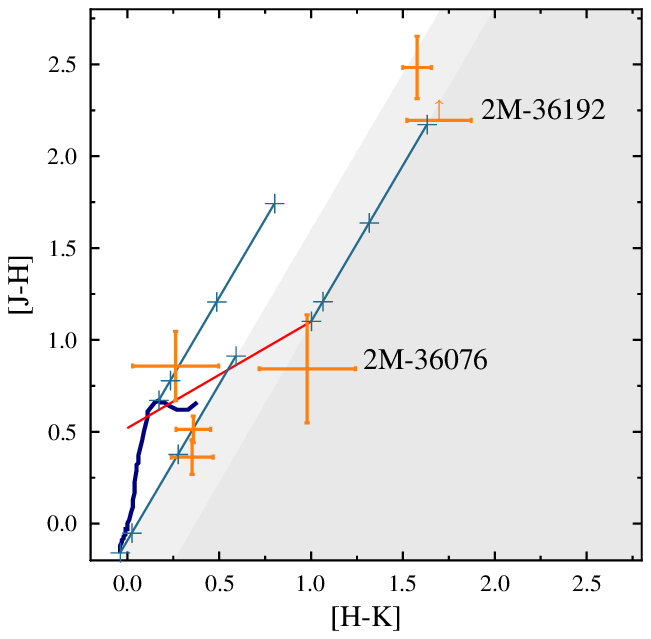}
   \caption{MMS19 field of view (center [J2000]: 08:49:08.5, -43:35:43.7) and colour-colour diagram.}
              \label{Fig:MMS19}
\end{figure*}
\clearpage

\begin{figure*}
   \centering
   \includegraphics[width=8cm]{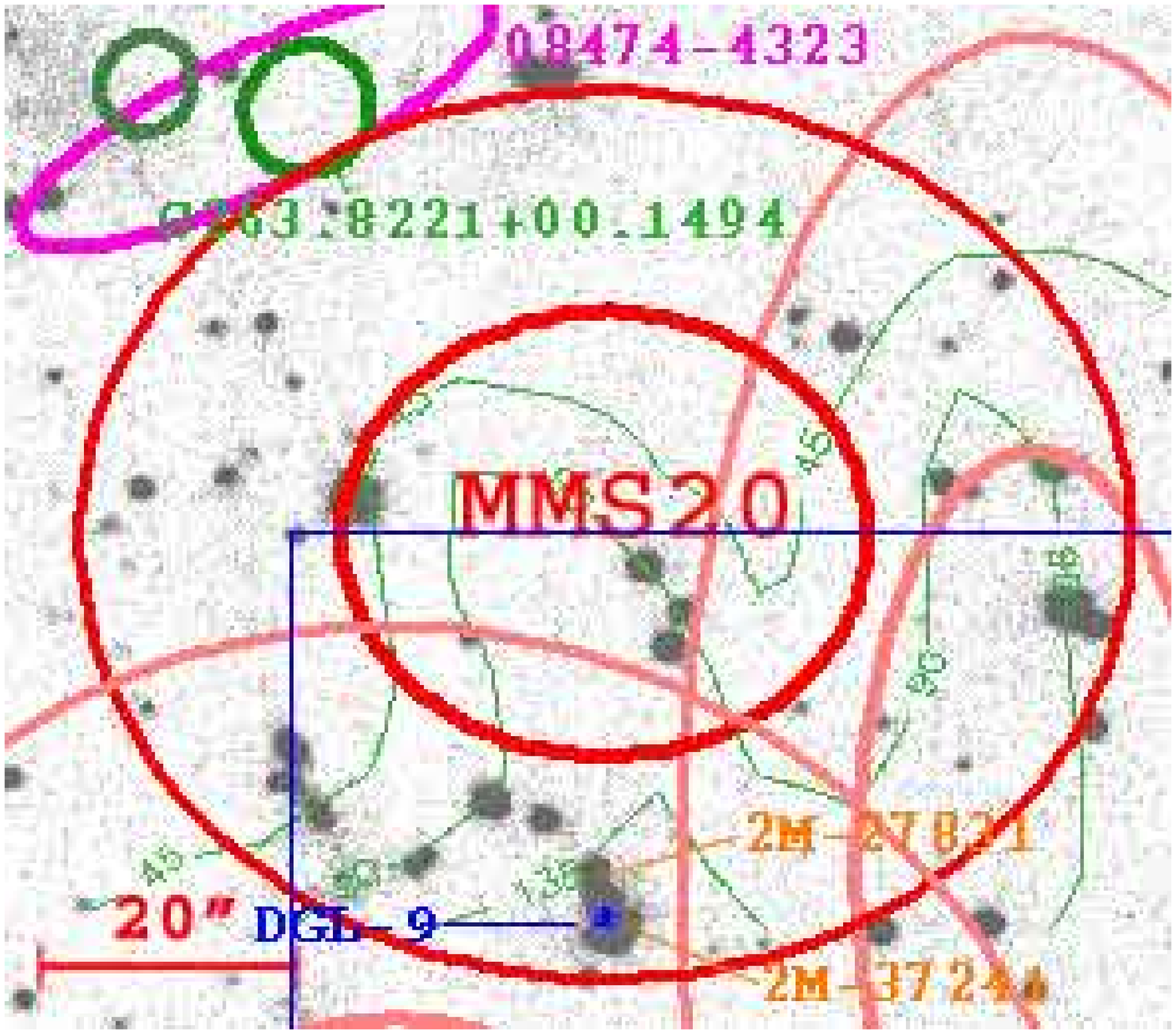}
\includegraphics[width=8cm]{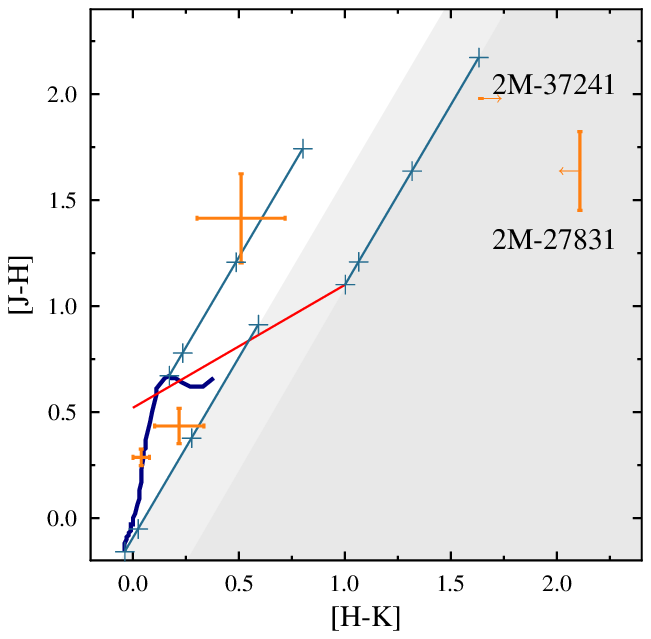}
   \caption{MMS20 field of view (center [J2000]: 08:49:11.2, -43:35:25.9) and colour-colour diagram.}
              \label{Fig:MMS20}
\end{figure*}

\begin{figure*}
   \centering
   \includegraphics[width=8cm]{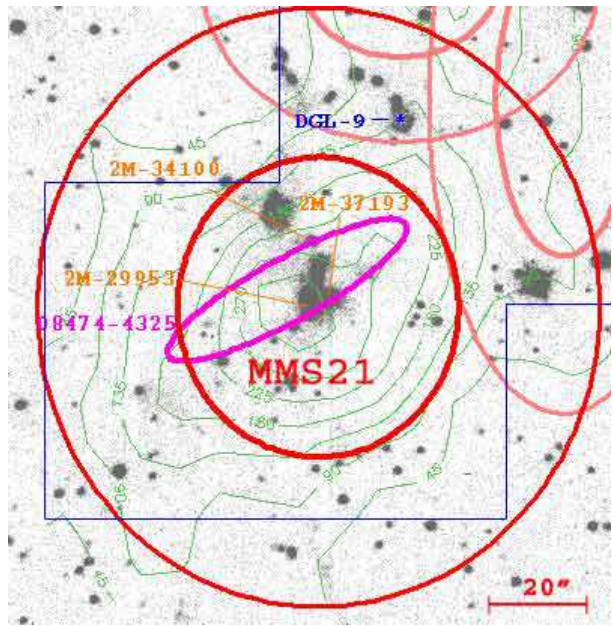}
   \caption{MMS21 field of view (center [J2000]: 08:49:13.0, -43:36:21.8).}
              \label{Fig:MMS21}
\end{figure*}
\begin{figure*}
   \centering
   \includegraphics[width=\textwidth]{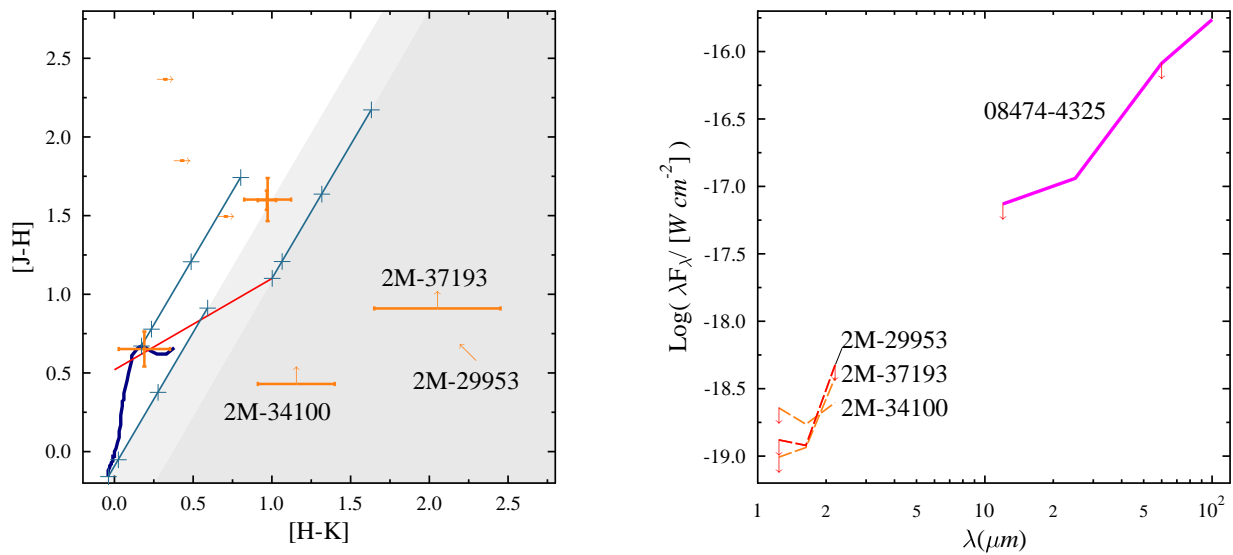}
   \caption{MMS21 colour-colour and SED diagrams.}
              \label{Fig:MMS21-colcol-SED}
\end{figure*}

\begin{figure*}
   \centering
   \includegraphics[width=8cm]{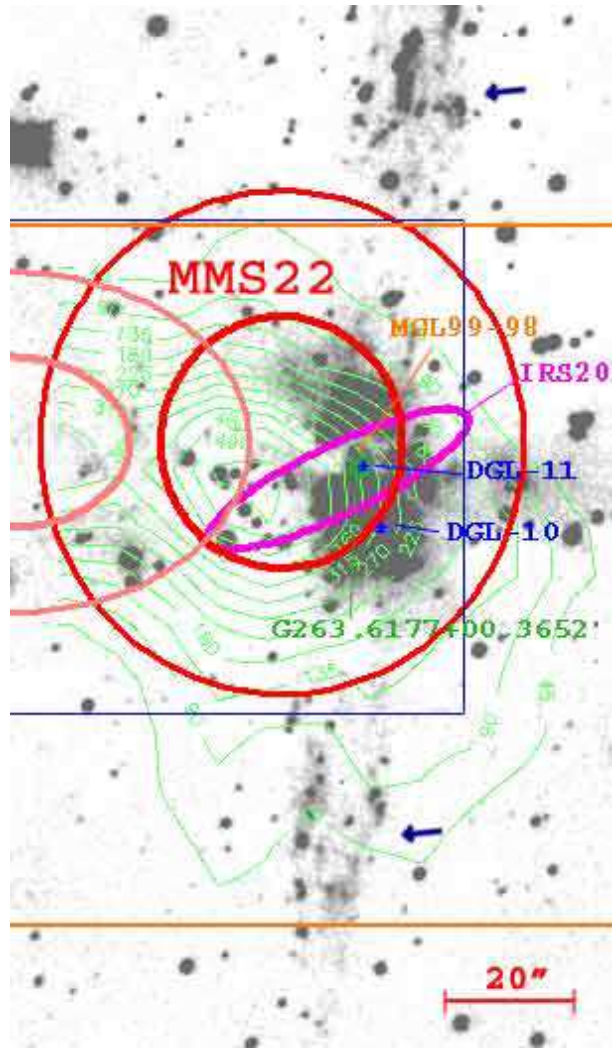}
   \caption{MMS22 field of view (center [J2000]: 08:49:26.0, -43:17:13.0).}
              \label{Fig:MMS22}
\end{figure*}
\begin{figure*}
   \centering
   \includegraphics[width=\textwidth]{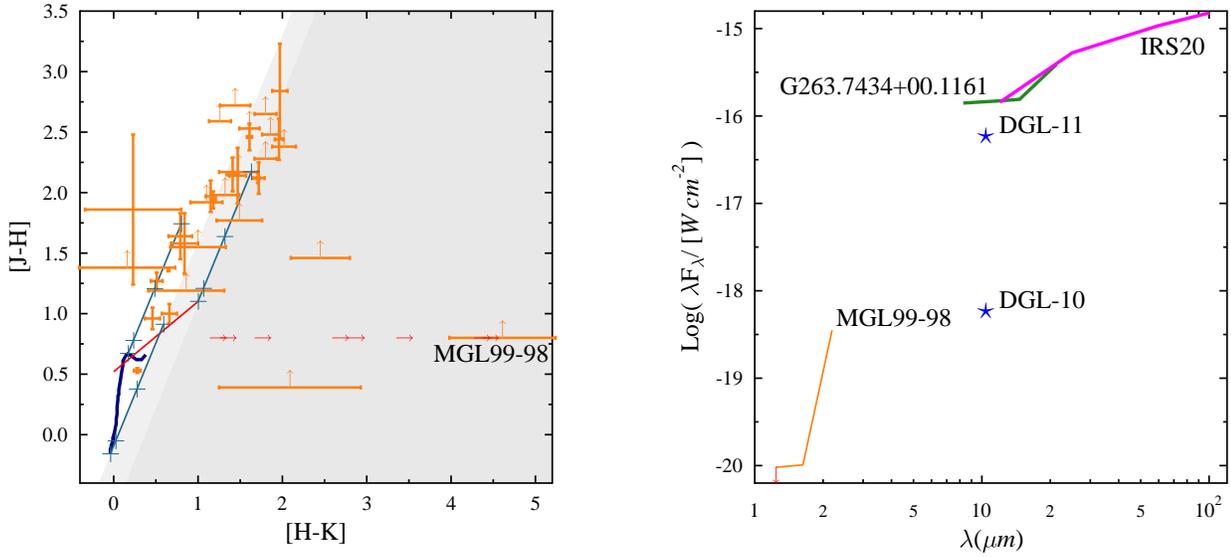}
   \caption{MMS22 colour-colour and SED diagrams.}
              \label{Fig:MMS22-colcol-SED}
\end{figure*}

\begin{figure*}
   \centering
   \includegraphics[width=5cm]{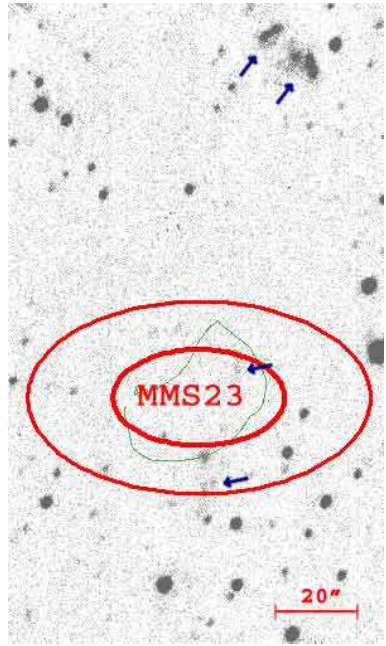}
   \caption{MMS23 field of view (center [J2000]: 08:49:30.2, -44:04:10.0).}
              \label{Fig:MMS23}
\end{figure*}

\begin{figure*}
   \centering
   \includegraphics[width=6cm]{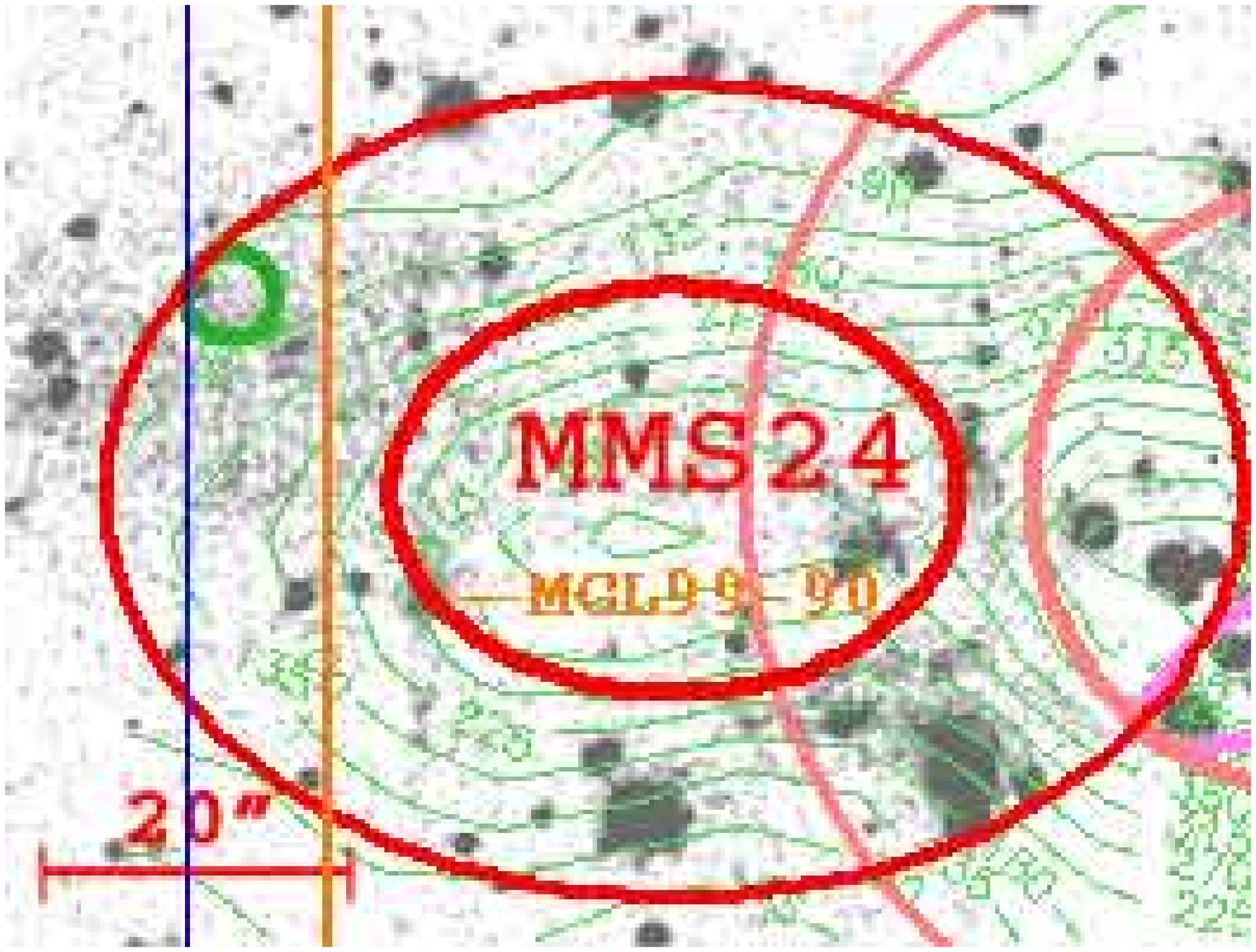}
\includegraphics[width=8cm]{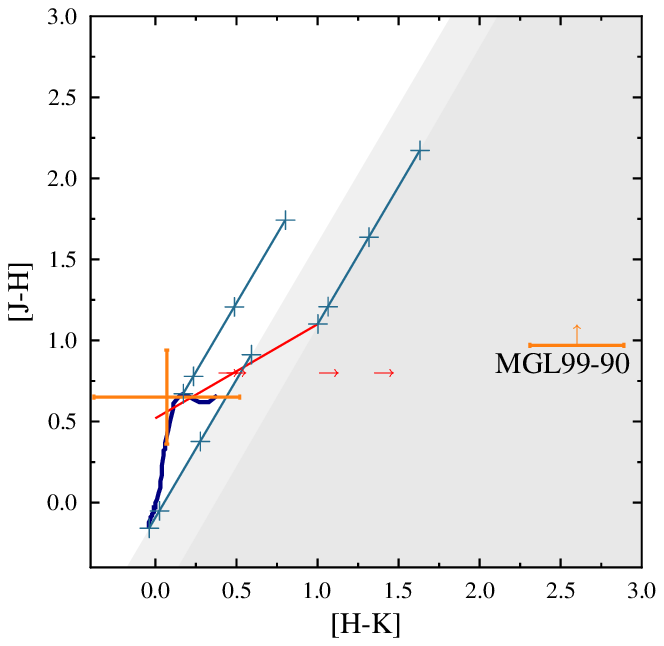}
   \caption{MMS24 field of view (center [J2000]: 08:49:30.1, -43:17:00.2) and colour-colour diagram.}
              \label{Fig:MMS24}
\end{figure*}

\begin{figure*}
   \centering
   \includegraphics[width=8cm]{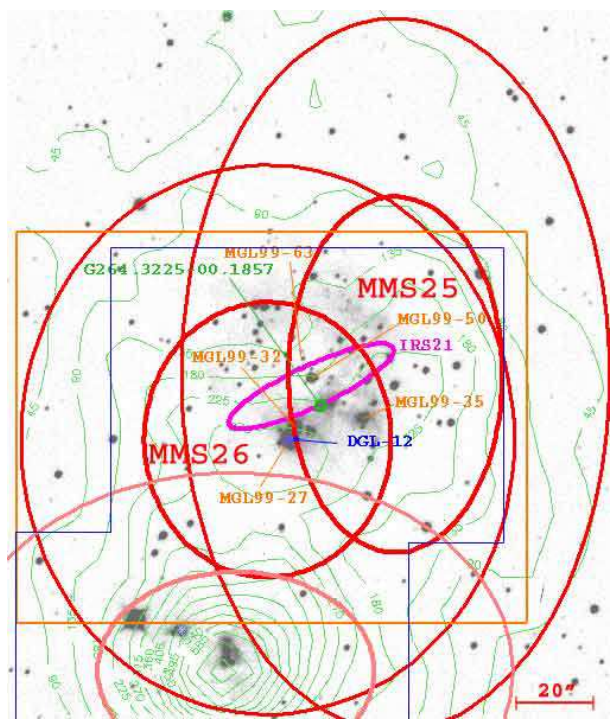}
   \caption{MMS25-26 field of view (center [J2000]: 08:49:33.5, -44:10:34.5).}
              \label{Fig:MMS25-26}
\end{figure*}
\begin{figure*}
   \centering
   \includegraphics[width=\textwidth]{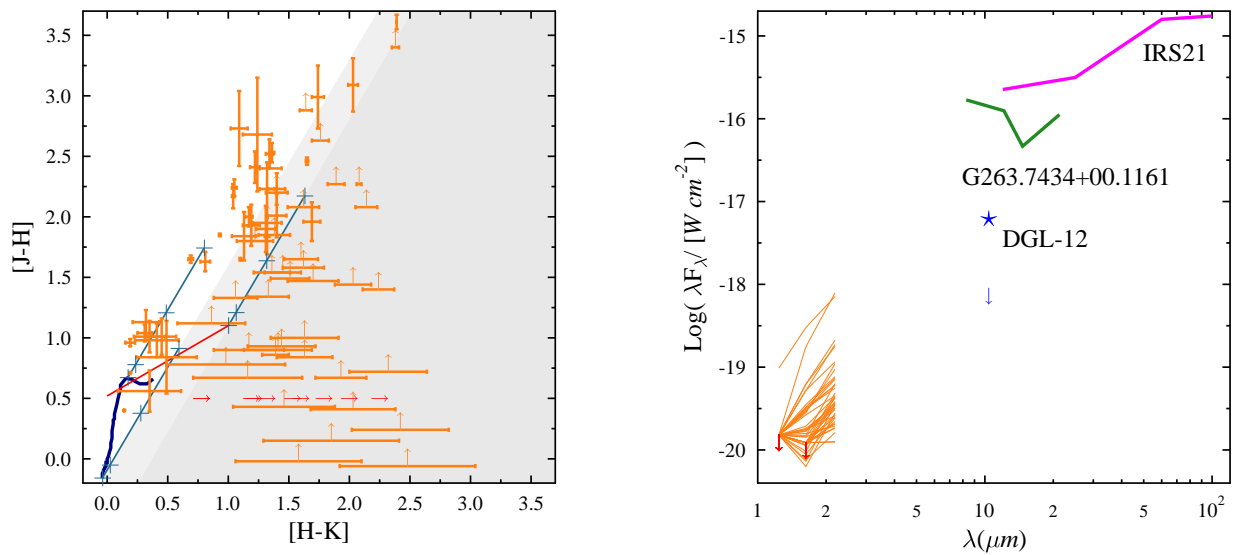}
   \caption{MMS25-26 colour-colour and SED diagrams.}
              \label{Fig:MMS25-26-colcol-SED}
\end{figure*}

\begin{figure*}
   \centering
   \includegraphics[width=10cm]{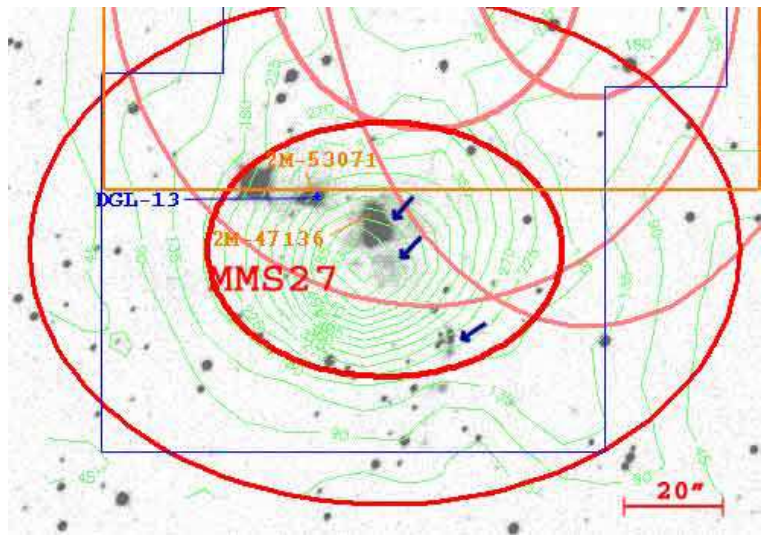}
\includegraphics[width=8cm]{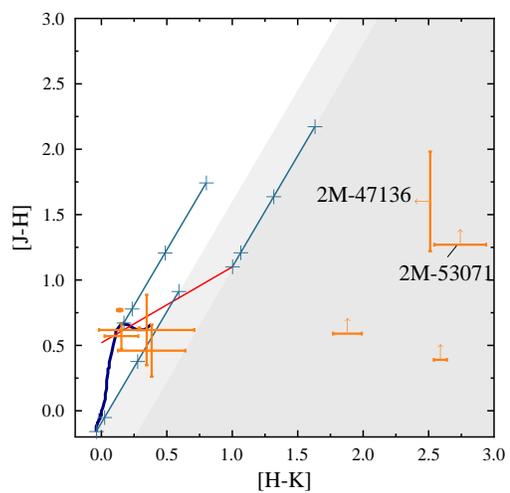}
   \caption{MMS27 field of view (center [J2000]: 08:49:35.2, -44:11:52.8) and colour-colour diagram.}
              \label{Fig:MMS27}
\end{figure*}

\begin{figure*}
   \centering
   \includegraphics[width=8cm]{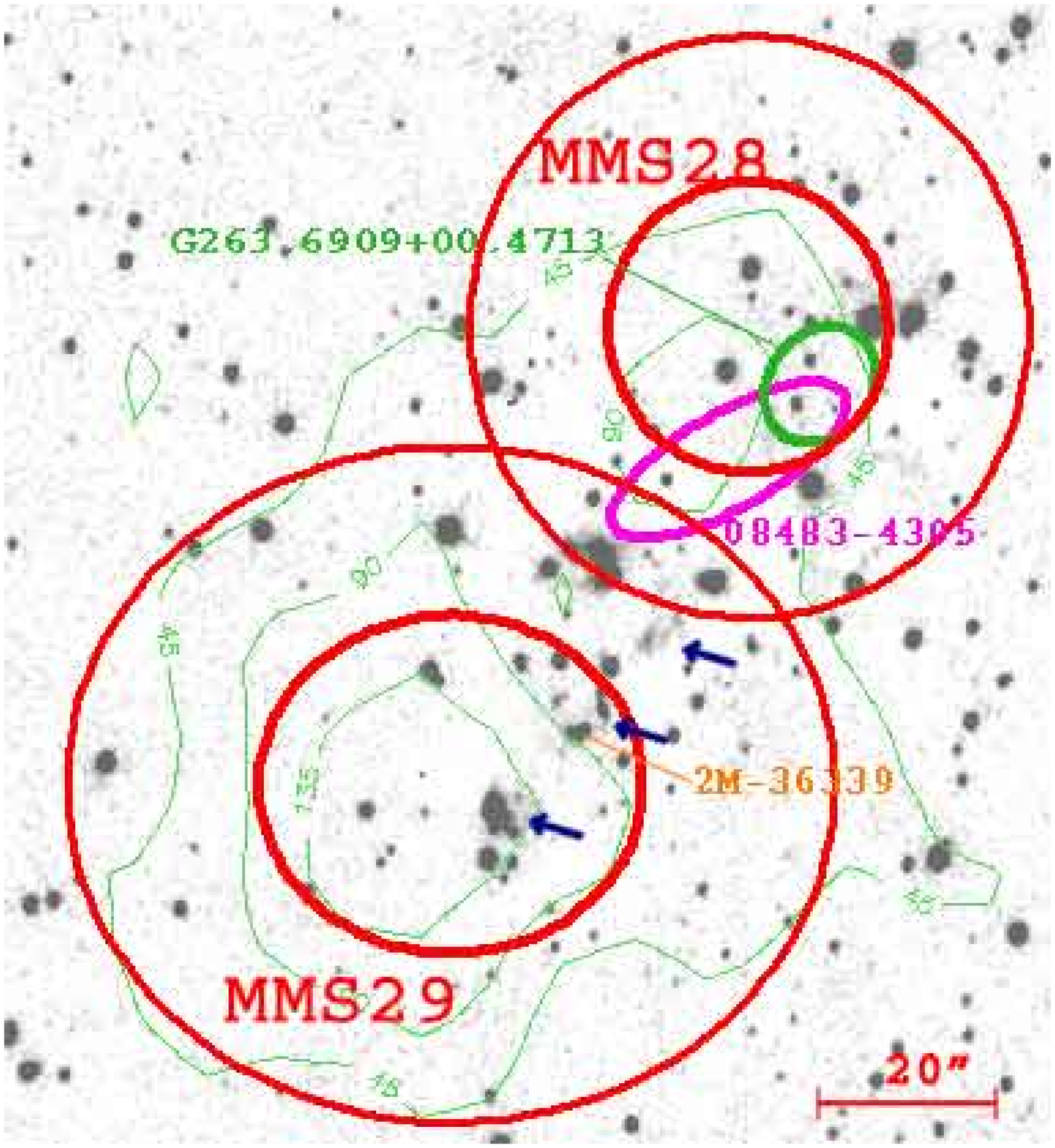}
\includegraphics[width=8cm]{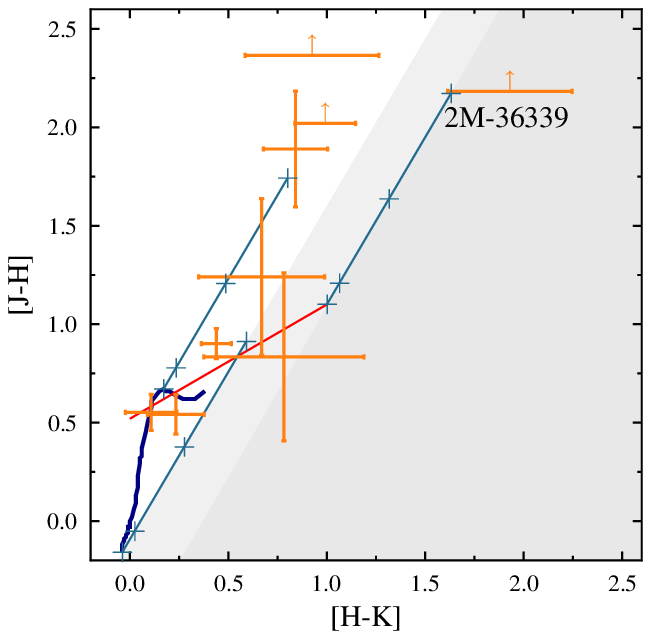}
   \caption{MMS28-29 field of view (center [J2000]: 08:50:10.0, -43:16:41.3) and colour-colour diagram.}
              \label{Fig:MMS28-29}
\end{figure*}

\clearpage


\clearpage

\end{document}